\documentclass[aps,prb,amssymb,twocolumn,amsfonts]{revtex4}
\usepackage{tabularx,graphicx,float}

\newcommand{\bk}{{\bf k}}

\newcommand{\bq}{{\bf q}}

\newcommand{\bK}{{\bf K}}

\newcommand{\br}{{\bf r}}

\newcommand{\bR}{{\bf R}}

\newcommand{\ef}{\epsilon_{\rm F}}

\newcommand{\vf}{v_{\rm F}}
\newcommand{\op}{\omega^{\prime}}

\newcommand{\sgn}{{\rm sgn}}
\newcommand{\xn}{\xi_n}
\newcommand{\xnp}{\xi_{n^{\prime}}}
\newcommand{\la}{\lambda}
\newcommand{\lap}{\lambda^{\prime}}
\newcommand{\np}{n_{>}}
\newcommand{\nm}{n_{<}}

\usepackage{graphics,graphicx,amsmath}
\usepackage{tabularx,graphicx,float}
\usepackage{epsfig}

\usepackage{subfigure}

\usepackage{bm}
\usepackage{wasysym}

\usepackage{ulem}

\usepackage{bm}
\usepackage{color}
\usepackage{verbatim}

\begin{document}

\bibliographystyle{apsrev}

\date{\today}

\author{R. Rold\'{a}n, M.O. Goerbig and J.-N. Fuchs}

\affiliation{Laboratoire de Physique des Solides, Univ. Paris-Sud,\\
CNRS, UMR 8502, F-91405 Orsay Cedex, France.\\}

\title{The magnetic field particle-hole excitation spectrum\\
in doped graphene and in a standard two-dimensional electron gas}

\begin{abstract}
The particle-hole excitation spectrum for doped graphene is calculated from the dynamical polarizability.
We study the zero and finite magnetic field cases and compare them to the standard two-dimensional electron gas.
The effects of electron-electron interaction are included within the random phase approximation. From the obtained polarizability, we study the screening effects and the collective
excitations (plasmon, magneto-excitons, upper-hybrid mode and linear magneto-plasmons). We stress the
differences with the usual 2DEG.
\end{abstract}

\maketitle

\section{Introduction}

The particle-hole excitation spectrum (PHES) of a material is an
extremely useful ingredient for the understanding of its electronic
properties, namely at low energies. It reveals, for example,
collective modes such as the plasmon, the electric polarizability of
the material, and the screening properties of the electrons in it.
In this paper, we review the particular PHES of two-dimensional
electron systems, both for non-relativistic electrons in a usual
two-dimensional electron gas (2DEG)\cite{AFS82} and for massless
electrons in graphene.\cite{CG09} The main focus of the present
review article is on the PHES and the collective modes of doped
graphene in a strong perpendicular magnetic field in the integer
quantum Hall regime.\cite{IWFB07,S07,BM08,TS08,BGL08,RFG09,FDR09} For pedagogical reasons, we compare them to the
corresponding results of the 2DEG\cite{GV05,KH84} as well as to the
PHES in graphene without a magnetic field, which has been
theoretically studied in detail recently.\cite{S86,A06,WSSG06,HS07}

The main difference between the PHES in graphene and that in the
2DEG, in a strong magnetic field $B$, stems from the quantization of
the electrons' kinetic energy in Landau levels (LL):  these levels
are equidistantly spaced in the case of non-relativistic electrons
in the 2DEG, $\epsilon_n^{\rm 2DEG}=\omega_c(n+1/2)$, in terms of
the cyclotron frequency $\omega_c=eB/m_b$, where $m_b$ is the band
mass (we use a system of units such that $\hbar \equiv 1$). In
contrast to this, the LLs in graphene occur in two copies as a
consequence of the two energy bands ($\lambda=+$ for the conduction
and $\lambda=-$ for the valence band), and their level spacing
decreases with increasing LL quantum number $n$,

\begin{equation}\label{relLL}
\epsilon_{\lambda,n}=\lambda\epsilon_n=\lambda
\frac{\vf}{l_B}\sqrt{2n}\propto \lambda\sqrt{Bn}\, .
\end{equation}
where $\vf$ is the Fermi velocity in graphene and $l_B=1/\sqrt{eB}$
is the magnetic length. This difference in LL quantization as well
as the absence of backscattering due to the chirality properties of
electrons in graphene \cite{SA98} yield a strikingly different PHES
for graphene when compared to the 2DEG. Whereas, in the latter, the
collective excitations are dominated by essentially horizontal
weakly-dispersing magneto-excitons,\cite{KH84} in addition to the
upper hybrid mode, those in graphene are linear magneto-plasmons
\cite{RFG09} that disperse roughly parallel to an energy line
$\omega=\vf q$, as a function of the wave vector $q$. The precursors
of these modes are already visible in the PHES for non-interacting
electrons and acquire coherence once electron-electron interactions
are taken into account, e.g. on the level of the random-phase
approximation (RPA).

The paper is organized as follows. In Sec. II, we introduce the
basic expressions for the polarization function of graphene in a
strong magnetic field. The intermediate steps of the derivation may
be found in Appendix A. Section III is devoted to a discussion of
the PHES in the standard 2DEG for non-relativistic electrons,
whereas that for doped graphene is presented in Sec. IV. Both
sections comprise also a short review of the $B=0$ case. In Sec. V,
we aim at a physical interpretation of main features of the two
different PHES in a strong magnetic field within a wave-function
analysis, and the screening properties in the static limit are
reviewed in Sec. VI.

\section{Polarizability}\label{SecCalculation}

The Hamiltonian for graphene in a magnetic field can be expressed as
${\cal H}_0^{K}=\vf\, {\pmb \pi}\cdot {\pmb \sigma}$ for the $K$
valley and ${\cal H}_0^{K'}=\vf\, {\pmb \pi}\cdot {\pmb \sigma^*}$
for the $K'$ valley, where ${\pmb \sigma=(\sigma^x,\sigma^y)}$ are
Pauli matrices, and ${\pmb \pi}={\bf p}+e{\bf A}$ is the
gauge-invariant momentum with  ${\bf p}=-i{\pmb \nabla}$, and $\bf
A$ is the vector potential. In the symmetric gauge, the latter reads
$ {\bf A}=(-By/2,Bx/2,0)$, where $B$ is the modulus of the magnetic
field that we choose in the $z$-direction. The Fermi velocity
$\vf=3ta/2$ is expressed in terms of the nearest-neighbor hopping
integral $t\simeq 3$ eV and the carbon-carbon distance $a\simeq 1.4$
\AA. The eigenstates components
$|\psi_{\zeta\alpha;\lambda,n,\ell}\rangle$ are:

\begin{align}\label{psi}
|\psi_{+A;\lambda, n,\ell}\rangle=  |\psi_{-B;\lambda, n,\ell} \rangle & = -i\lambda 1^*_n |n-1,\ell\rangle  \nonumber\\
 |\psi_{-A;\lambda, n,\ell}\rangle = |\psi_{+B;\lambda, n,\ell}\rangle & =  2^*_n |n,\ell\rangle
\end{align}
where $n$ is a positive integer, $\lambda=\pm$ for states of
positive/negative energy and $\lambda=0$ for $n=0$. The
corresponding eigenenergies are given in Eq. (\ref{relLL}). The
index $\zeta=+(-)$ denotes electrons in the $ {\rm K} ( {{\rm
K}^{\prime}})$ valleys and $\alpha=A(B)$ the $A(B)$ sublattice
component of the electronic wave function. We have furthermore
introduced the simplified notation $1^*_n=\sqrt{(1-\delta_{n,0})/2}$
and $2^*_n=\sqrt{(1+\delta_{n,0})/2}$. Here the quantum number $n$
labels the LL, whereas the other quantum number $\ell$, which
determines the LL degeneracy, varies from  $0$ to $N_B-1$, with
$N_B={\cal
A}n_B={\cal A}/2\pi l_B^2$, in terms of the total sample surface $\cal A$ . 
The states $|n,\ell\rangle $ are the eigenvectors of the Hamiltonian
\begin{equation}\label{nonrelH}
 \mathcal{H}_0^{\rm 2DEG}=\frac{\pi_x^2 + \pi_y^2}{2m_b}
\end{equation}
for the standard 2DEG in a magnetic field.

The bare polarization function
\begin{equation}\label{Pi0}
i\Pi^0(\bq,\omega)=\int \frac{d\op d\bk}{(2\pi)^3}\,{\rm Tr}\left[
G^0(\bk,\op)G^0(\bk+\bq,\op+\omega)\right ].
\end{equation}
may be calculated with the help of the Green's functions
$G^0(\bk,\omega)$ for non-interacting electrons [see Eq.
(\ref{G0komega})] and reads, for the case of a strong magnetic
field,\cite{RFG09}
\begin{equation}\label{Pi+}
\Pi^0(\bq,\omega)=\sum_{\la\lap}\sum_{n,n^{\prime}}\frac{\Theta(\lambda^{\prime}\xnp)-\Theta(\lambda\xn)}{\lambda\xn-\lambda^{\prime}\xnp
+\omega+i\delta}{\cal
\overline{F}}_{nn^{\prime}}^{\lambda\lambda^{\prime}}(\bq).
\end{equation}
The expression of the functions ${\cal \overline{F}}_{nn^{\prime}}^{\lambda\lambda^{\prime}}(\bq)$ and the details of the calculation
may be found in Appendix \ref{App:Pol}. $\Pi^0$ contains two separate contributions,

\begin{equation}\label{Pi0Graph}
\Pi^0(\bq,\omega)=\sum_{n=1}^{N_F}\Pi_{n}^{\la_F}(\bq,\omega)+\Pi^{vac}(\bq,\omega).
\end{equation}
The vacuum contribution $\Pi^{vac}(\bq,\omega)$ takes into account
inter-band processes, whereas
$\sum_{n=1}^{N_F}\Pi_{n}^{\la_F}(\bq,\omega)$ represents intra-band
contributions when the Fermi energy $\epsilon_F=\epsilon_{N_F}$ lies
in the conduction band, as we assume implicitly from now on.

\subsection{Effect of electron-electron interaction}

From $\Pi^0(\bq,\omega)$ we can calculate the renormalized
polarization function in the RPA, which is defined as
\begin{equation}\label{EqPiRPA}
\Pi^{RPA}(\bq,\omega)=\frac{\Pi^0(\bq,\omega)}{\varepsilon^{RPA}(\bq,\omega)}=\frac{\Pi^0(\bq,\omega)}{1-v(\bq)\Pi^0(\bq,\omega)}
\end{equation}
where $v(\bq)$ is the unscreened two-dimensional Coulomb potential
in momentum space
\begin{equation}
v(\bq)=\frac{2\pi e^2}{\varepsilon_b |\bq|}
\end{equation}
in terms of the background dielectric constant $\varepsilon_b$, and
$\varepsilon^{RPA}(\bq,\omega)=1-v(\bq)\Pi^0(\bq,\omega)$ is the
dielectric function. Long-range electron-electron interaction
usually leads to the appearance of collective modes in the spectrum,
such as the plasmon, the dispersion of which is defined from the
zeros of the dielectric function
\begin{equation}\label{Eq:Plasmon}
\varepsilon^{RPA}(\bq,\omega)=0.
\end{equation}
Collective modes will be discussed in detail for both, graphene and
a standard 2DEG, in the following sections.

\section{Particle-hole excitation spectrum of a standard 2DEG}

\begin{figure}[t]
\centering
  \subfigure[]{\label{PHES2DEGB0}\includegraphics[width=0.23\textwidth]{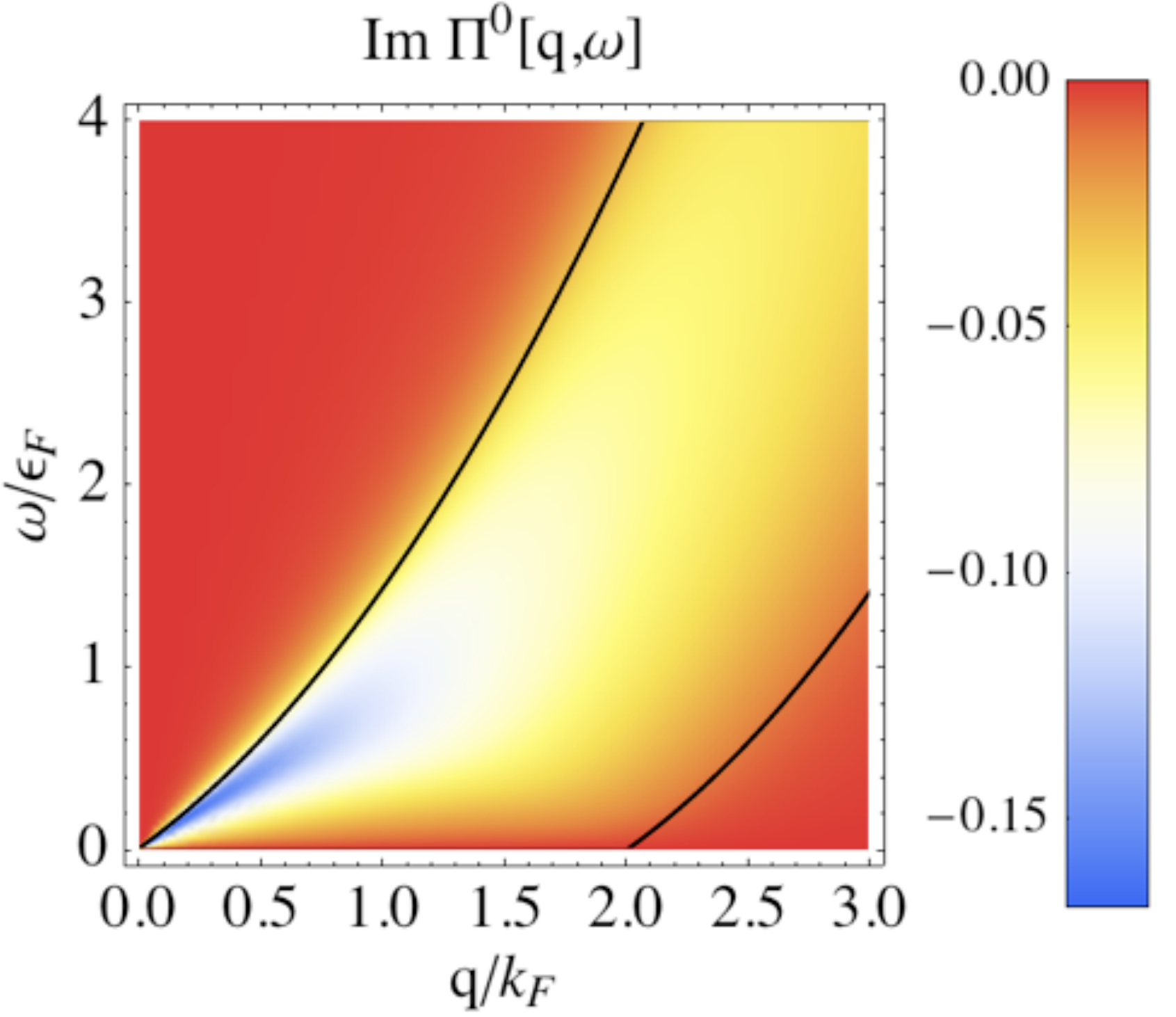}}
      \subfigure[]{\label{PHESRPA2DEGB0}\includegraphics[width=0.23\textwidth]{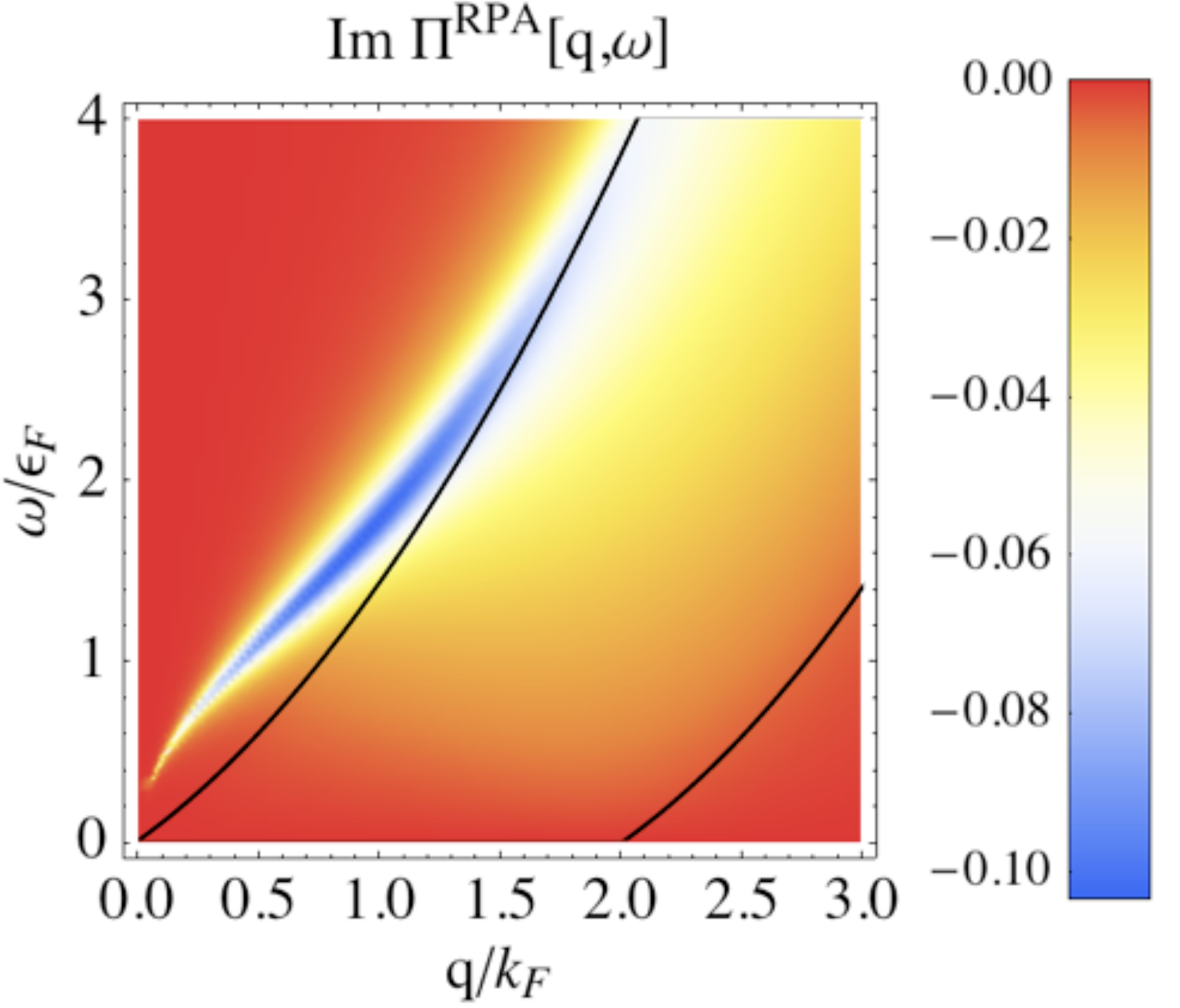}}
   \subfigure[]{\label{ImPi0DP2DEG}\includegraphics[width=0.23\textwidth]{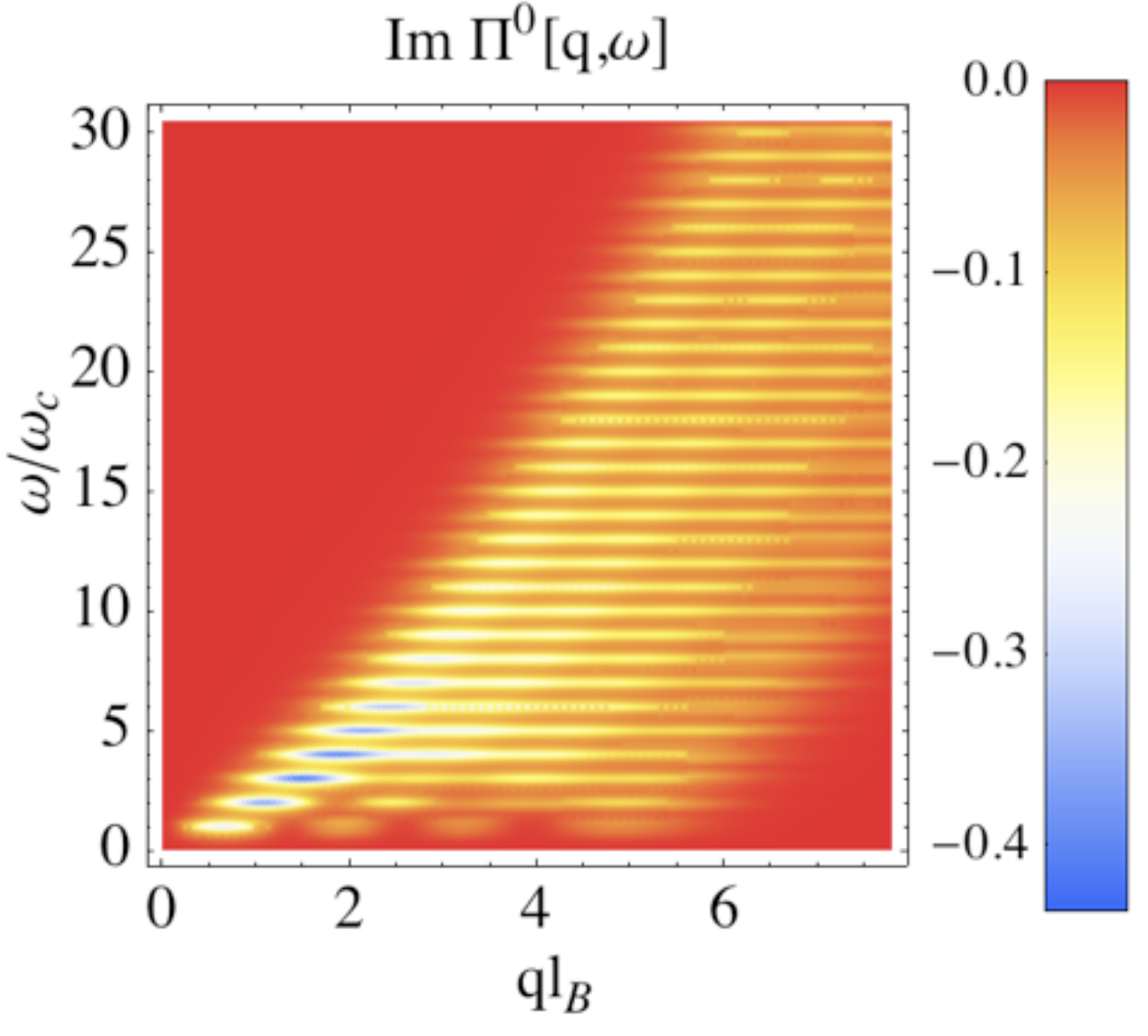}}
 \subfigure[]{\label{ImPiRPA2DEG}\includegraphics[width=0.23\textwidth]{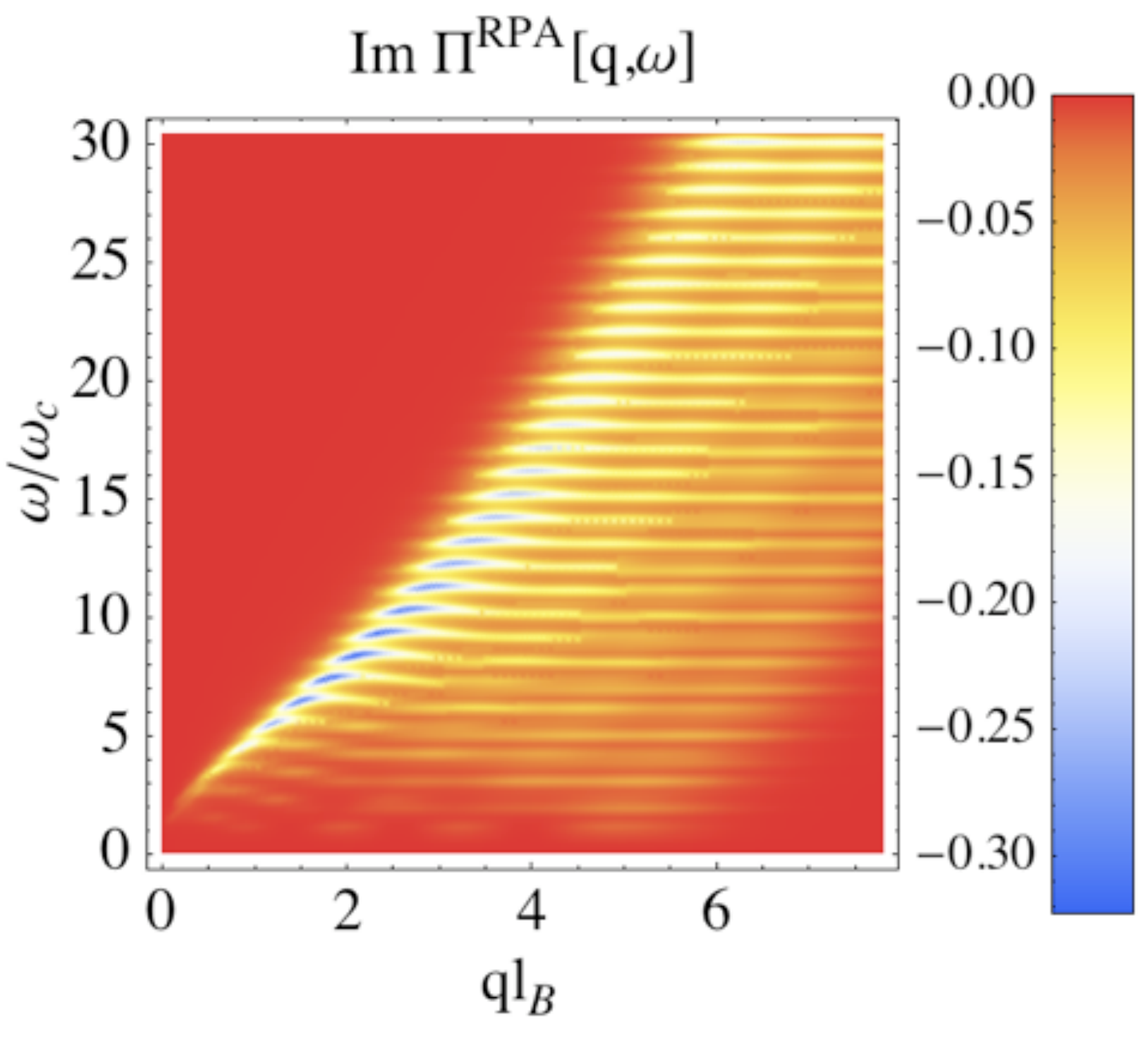}}
\caption{Density plot of ${\rm Im}\Pi(\bq,\omega)$. Plots (a)
corresponds to non-interacting polarization of a 2DEG, Eq.
(\ref{EqPi0B02DEG}), whereas (b) accounts also for electron-electron
interactions in the RPA. Plots (c) and (d) show the imaginary part
of the non-interacting and RPA polarization function, respectively,
of a 2DEG in a magnetic field. In (a) and (b) we have used
$\delta=0.2 \varepsilon_F$. In (c) and (d), $N_F=3$ and
$\delta=0.2\omega_c$. In (b) and (d), we have used $r_s\simeq 3$.}
  \label{PHESB02DEG}
\end{figure}

In this section we briefly review the results for the PHES in a 2DEG
and start with the case of zero magnetic field. The polarization
function Eq. (\ref{Pi0}) for a system of free electrons with
parabolic band $\varepsilon_{\bk}=k^2/2m_b$ can be expressed, after
some manipulation, as \cite{GV05}

\begin{eqnarray}\label{EqPi0B02DEG}
\Pi^0_{\rm 2DEG}(\bq,\omega)=\rho(\epsilon_F)\frac{k_F}{q}\!\!\!\!&\!\!\!\!&\!\!\!\!\left [ \Psi \left ( \frac{\omega +i\delta}{\vf q}-\frac{q}{2k_F}\right )\right.\nonumber\\
&&\left. -\Psi \left ( \frac{\omega +i\delta}{\vf q}+\frac{q}{2k_F}\right ) \right ]
\end{eqnarray}
where $\rho(\epsilon_F)$ is the density of states at the Fermi
level, $\delta\sim \tau^{-1}$ accounts for a finite life time of the
quasiparticles, and $\Psi(z)=z-\sgn ({\rm Re}\,z)\sqrt{z^2-1}$. In
Fig. \ref{PHESB02DEG}(a) we show a density plot of ${\rm Im}\,\Pi^0_{\rm 2DEG}$. From the imaginary part of $\Pi^0$ we can obtain the
PHES, which is the region of the momentum-energy plane where it is
possible to excite electron-hole pairs. This region is defined, for
a non-interacting electron gas in the absence of a magnetic field,
as the continuum of Fig. \ref{PHES2DEGB0} with non-zero ${\rm
Im}\,\Pi^0$, which corresponds (for $\delta\rightarrow 0$) to the
region delimited by the solid black lines. The boundaries of the
spectrum are defined by $\max[0,\omega_-(q)]\le |\omega|\le
\omega_+(q)$, where $\omega_{\pm}(q)=q^2/2m_b\pm \vf q$, in terms of
the Fermi velocity $\vf=\partial\epsilon/\partial
k|_{k_F}=k_F/m_b=\sqrt{2\epsilon_F/m_b}$, which in contrast to
graphene depends on $\epsilon_F$. Notice that this PHES is not
uniform, but presents some structure. Apart from the
quasi-homogeneous yellow region, two other zones are worth
describing: the blue one, with a strong spectral weight, which is
the precursor of the plasmon mode, as we will see below, and the
almost red low energy region, with a very weak spectral weight and
which, in a 1D system, would belong to the forbidden zone of the spectrum for
particle-hole excitations.

Electron-electron interactions lead to the appearance of a
collective mode, the plasmon, which can be captured within the RPA.
The imaginary part of $\Pi^{RPA}(\bq,\omega)$, shown in Fig.
\ref{PHESRPA2DEGB0}, reveals above the continuum boundaries a
well-defined peak centered at $\omega=\omega_p$, the frequency of
the plasmon mode. The broadening of the peak depends on the value of
$\delta\sim \tau^{-1}$ and is due, e. g., to scattering of
electrons by disorder. The dispersion relation of the plasmon may be
calculated from the zeros of the RPA polarization function. From the
$q\rightarrow 0$ and $\omega\ll \vf q$ expansion of the polarization
function, the long wavelength limit of the plasmon dispersion is
found to be\cite{S67}
\begin{equation}\label{EqPlasmonB02DEG}
\omega_p(q)\simeq\sqrt{\frac{2e^2\ef}{\varepsilon_b}q+\frac{3}{4}\vf^2
q^2},
\end{equation}
where the Fermi energy in a 2DEG with a parabolic band can be
expressed in terms of the the uniform density of electrons $n_{el}$
as $\ef=\pi n_{el}/m_b$. Furthermore, from the analytic solution of
Eq. (\ref{Eq:Plasmon}), an exact dispersion relation of the plasmon
mode may be obtained to all orders in $q$ (see Ref.
\onlinecite{CHSS82}). At low energies, it disperses as $\sqrt{q}$,
and at some critical wave vector $q_c$, the mode touches the
electron-hole continuum. This critical value may be obtained from
\cite{CHSS82}
\begin{equation}\label{Eq:qc2DEG}
\frac{(q_c/k_F)^2 }{\sqrt{2}r_s}+\frac{(q_c/k_F)^3}{4r_s^2}=1,
\end{equation}
in terms of the dimensionless interaction parameter  $r_s\equiv 2m_be^2/\varepsilon_bk_F$. Above $q_c$, the plasmon disperses
roughly parallel to the boundary of the particle-hole continuum while being Landau-damped due to its decay into
electron-hole pairs [see Fig. \ref{PHESRPA2DEGB0}].

In the presence of a strong magnetic field perpendicular to the 2DEG, the bare polarizability can be expressed as\cite{KH84}
\begin{equation}\label{EqPi02DEG}
\Pi^0_{2DEG}(\bq,\omega)=\sum_{m=1}^{N_c}{\sum}'\frac{{\cal
F}_{n,m}(\bq)}{\omega-m\omega_c+i\delta}+(\omega^+\rightarrow
-\omega^-)
\end{equation}
where $\sum'=\sum_{n=\max(0,N_F-m)}^{N_F-1}$ and
$\omega^+\rightarrow -\omega^-$ indicates the replacement
$\omega+i\delta\rightarrow -\omega-i\delta$. The form factor due to
the wave-function overlap is
\begin{equation}\label{FF2DEG}
{\cal F}_{n,m}(\bq)=\frac{e^{-\frac{q^2l_B^2}{2}}}{2\pi
l_B^2}\frac{n!}{(n+m)!}\left (\frac{q^2l_B^2}{2}\right)^m
\left[L_n^m\left(\frac{q^2l_B^2}{2}\right )\right]^2.
\end{equation}
A density plot of ${\rm Im}\,\Pi^0$ is shown in Fig. \ref{ImPi0DP2DEG} for
$N_F=3$. In the presence of a strong magnetic field, ${\rm
Im}\,\Pi^0(\bq,\omega)$ is a sum of Lorentzian peaks centered at
$\omega=m\omega_c$, with $m\equiv n'-n\ge 1$, the difference between
the LL indices of the electron $n'$ and the hole $n$. Therefore the
PHES is chopped into horizontal lines, separated by a constant
energy $\omega_c$. The width of each horizontal line is proportional
to the disorder broadening of the Landau levels. (The peaks become $\delta$-functions in
the clean limit $\delta\rightarrow 0^+$.) Notice that within each
horizontal line, the spectral weight is not homogeneously
distributed; one observes indeed a superstructure of $N_F+1$
brighter regions in Fig. \ref{ImPi0DP2DEG} following lines parallel
to the edges of what used to be the particle-hole continuum in the
zero-field case. These edges delimit, also for non-zero values of
the magnetic field, the PHES region of non-vanishing spectral
weight.

In the presence of electron-electron interactions, the density
fluctuation spectrum of a 2DEG at integer filling factors is
dominated by a single set of collective modes, known as horizontal
magneto-excitons. The frequency of these modes tend to $m\omega_c$
(where $m$ is a positive integer) in the long wavelength limit, and
their dispersion was calculated by Kallin and Halperin in the RPA
and in the time-dependent Hartree-Fock approximation.\cite{KH84} In
Fig. \ref{ImPiRPA2DEG} we show the RPA excitation spectrum for
$r_s\equiv 2m_b e^2/ \varepsilon_b k_F\approx 3$. Furthermore, one
notices that the plasmon energy is renormalized by the magnetic
field and evolves into the so called upper hybrid mode. Its
dispersion relation can be expressed as\cite{CQ74}
\begin{equation}\label{EqUH}
\omega_{uh}^2(q)=\omega_p^2(q)+\omega_c^2,
\end{equation}
where an approximate expression for $\omega_p(q)$ has been given in
Eq. (\ref{EqPlasmonB02DEG}).

\section{Particle-hole excitation spectrum of doped graphene}

\begin{figure}[t]
  \centering
   \subfigure[]{\label{ImPi0B0DP}\includegraphics[width=0.23\textwidth]{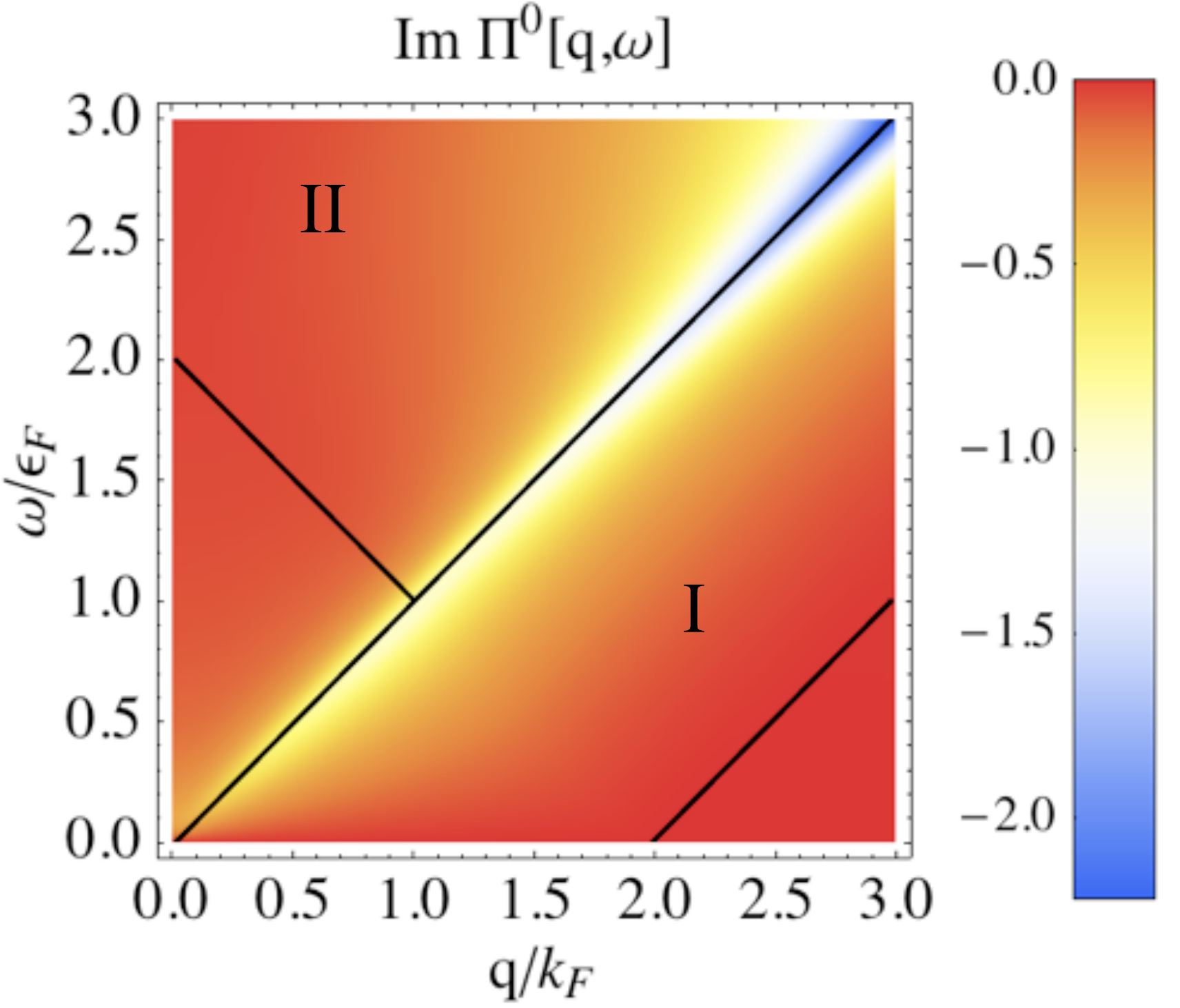}}
 \subfigure[]{\label{ImPiRPAB0DP}\includegraphics[width=0.23\textwidth]{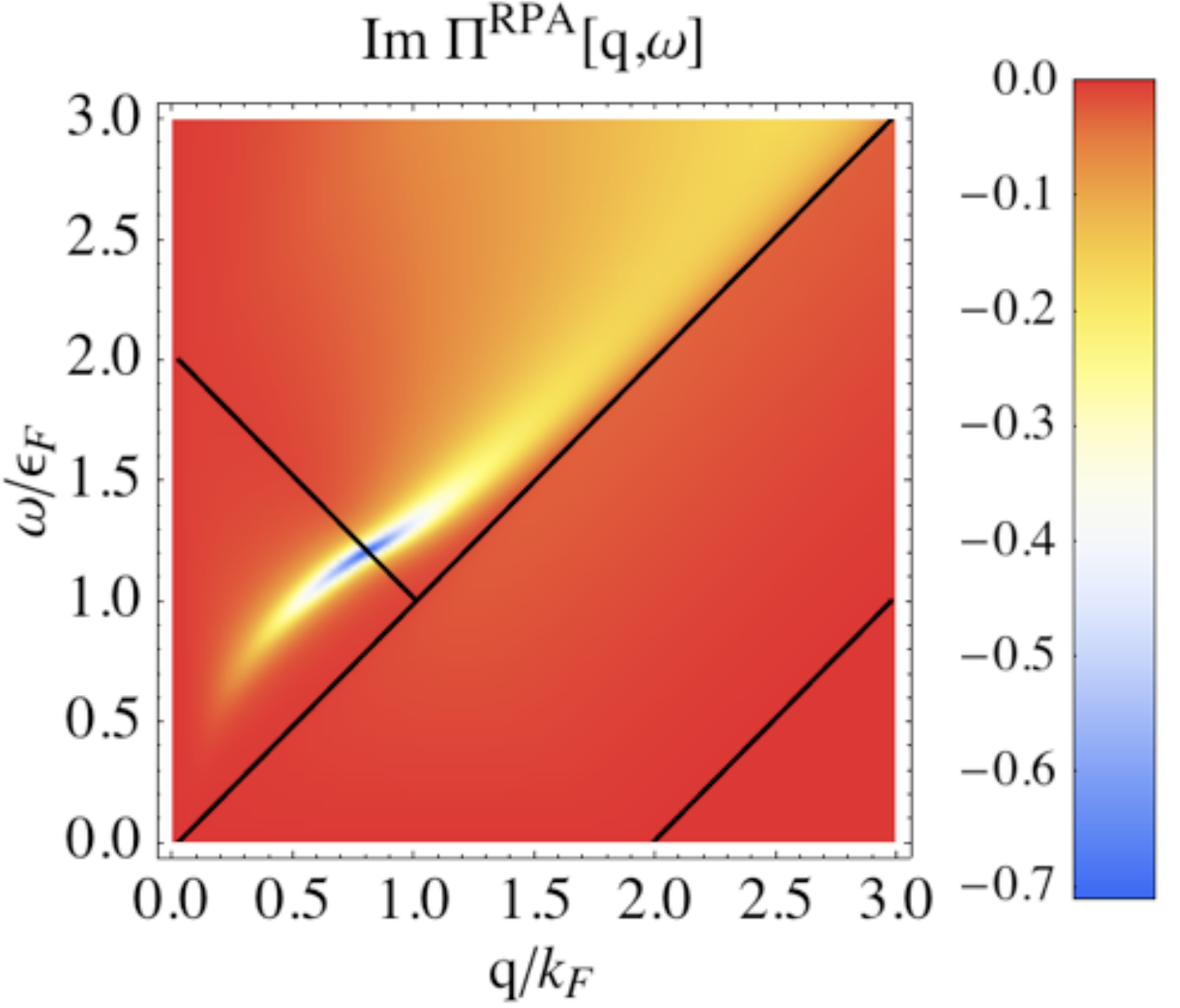}}
    \subfigure[]{\label{ImPi0DP}\includegraphics[width=0.23\textwidth]{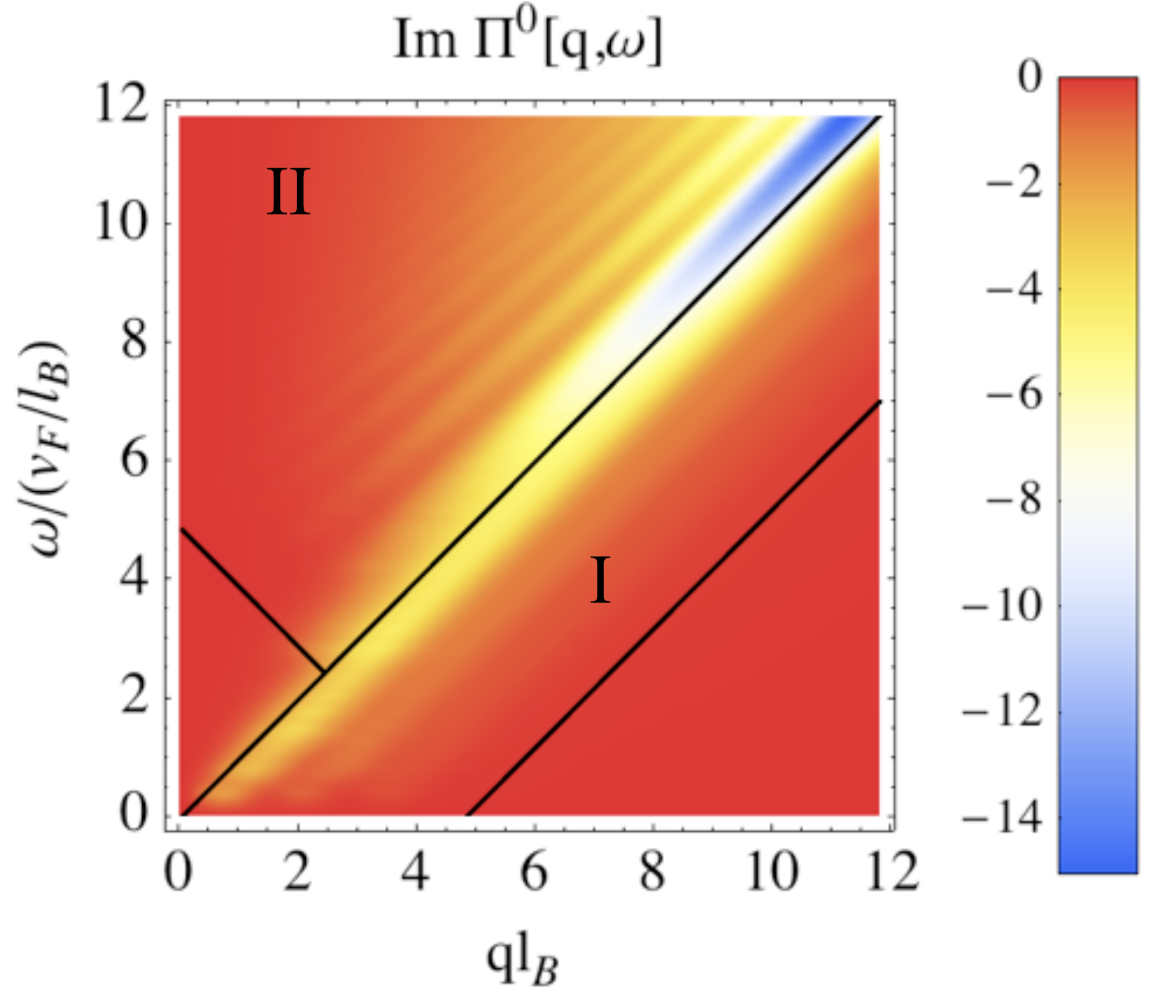}}
   \subfigure[]{\label{ImPiRPADP}\includegraphics[width=0.24\textwidth]{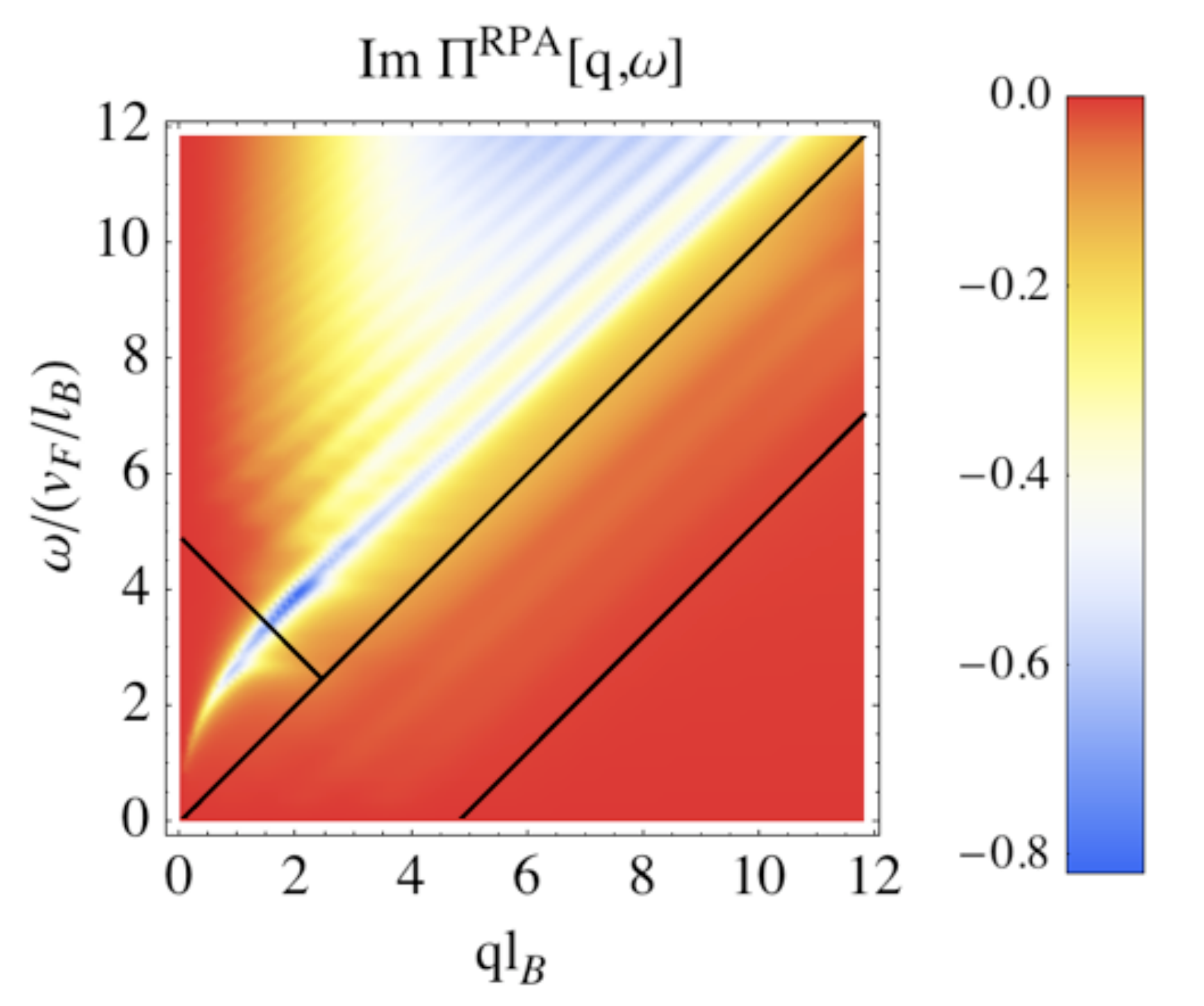}}
  \caption{Same as Fig. \ref{PHESB02DEG} but for the case of graphene.
  The solid lines delimitate the boundaries of the PHES. Region I corresponds to
  the intra-band and region II to the inter-band region of the PHES of doped graphene. In (a) and (b), $\delta=0.1 \varepsilon_F$. In (c) and (d), $\delta=0.2\vf/l_B$. In (b) and (d), we have used $r_s=1$.}
  \label{PiDinamicB0Graph}
\end{figure}

We start with a review of the polarization function in the absence
of a quantizing magnetic field, which turns out to be useful in the
understanding of the PHES also at $B\neq 0$. In the continuum approximation, the non-interacting
polarization function at zero temperature can be calculated from
\begin{equation}
\Pi^0(\bq,\omega)=\frac{g}{{\cal A}}\sum_{\bk,\lambda,\lambda'}\frac{\Theta(\lambda^{\prime}\xi_{\bk+\bq})-\Theta(\lambda\xi_{\bk})}{\lambda\xi_{\bk}-\lambda^{\prime}\xi_{\bk+\bq}
+\omega+i\delta}F_{\lambda\lambda'}(\bk,\bk+\bq),
\end{equation}
where $\lambda\xi_{\bk}=\lambda\vf |\bk|-\ef$ is the quasiparticle energy relative to the Fermi level, $g=g_sg_v=4$ accounts for spin and valley degeneracy and 
\begin{equation}\label{OverlapB0}
F_{\lambda\lambda'}(\bk,\bk+\bq)=\frac{1+\lambda\lambda'\cos\theta_{\bk,\bk+\bq}}{2}
\end{equation}
is the chirality factor or wave function overlap, where $\theta_{\bk,\bk+\bq}$ is the angle between $\bk$ and $\bk+\bq$. $\Pi^0(\bq,\omega)$ was calculated for undoped graphene in Ref. \onlinecite{GGV94}, where it was found that ${\rm
Im}\,\Pi^0(\bq,\omega)\sim q^2(\omega^2-\vf^2q^2)^{-1/2}$, which
implies that the {\it massless} non-interacting electrons in
graphene have an infinite response at the threshold $\omega=\vf q$.
The reason for this behavior is twofold: first the threshold is
determined by the linear dispersion relation and second, the
chirality factor (\ref{OverlapB0}) suppresses backscattering in graphene. This feature is still present in doped graphene,
although the form of the polarization function is richer than in the
absence of doping,\cite{S86,WSSG06,HS07} as may be seen in Fig.
\ref{PiDinamicB0Graph}, where we show a density plot of
$\Pi(\bq,\omega)$ for doped graphene at $B=0$. One notices that most
of the spectral weight is actually concentrated around $\omega=\vf
q$, as one would expect from the suppression of backscattering
($2k_F$ processes). This is similar to the case of zero doping.

In doped graphene, however, there are two regions of non-vanishing
spectral weight which arise from intra-band (region I) and
inter-band processes (region II). These two regions are separated by
the diagonal line $\omega=\vf q$. For zero doping, there are
naturally only inter-band processes. The intra-band contributions
are restricted to region I the boundaries of which are $\max(0,\vf
q-2\varepsilon_F)\le |\omega|\le \vf q$. This is the only kind of
processes present in the 2DEG, discussed in the preceding section.
The features of this zone are, apart from the different shape of the
boundaries in the two cases, similar to the PHES of the 2DEG,
although the spectral weight is no longer homogeneously distributed
within this region but vanishes when approaching the right boundary
as a consequence of the chirality factor. Inter-band particle-hole
excitations are restricted to region II in the PHES, with boundaries
$|\omega|\ge \max(\vf q,-\vf q+2\varepsilon_F)$. Indeed, the
inter-band excitation of lowest energy needs to overcome the Fermi
energy, which is therefore the lower bound of region II and which
approaches zero in  undoped graphene, where the region of inter-band
excitations covers the whole part $\omega>\vf q$  of the spectrum.
Furthermore, due to the presence of two energy bands, there is the
possibility of direct transitions from the valence to the conduction
band with $q=0$ momentum transfer. The $q=0$ transition of lowest
energy involves an energy cost that is twice the Fermi energy, which
yields a gapped region in the PHES in Fig.
\ref{PiDinamicB0Graph}(a), defined by $0<\omega<2\varepsilon_F$.
Notice however that the $q=0$ transition is suppressed due to the
chirality factor (\ref{OverlapB0}).

\begin{figure}[t]
  \centering
  \subfigure[]{\label{Pi0B02DEG-kF0_5}\includegraphics[width=0.23\textwidth]{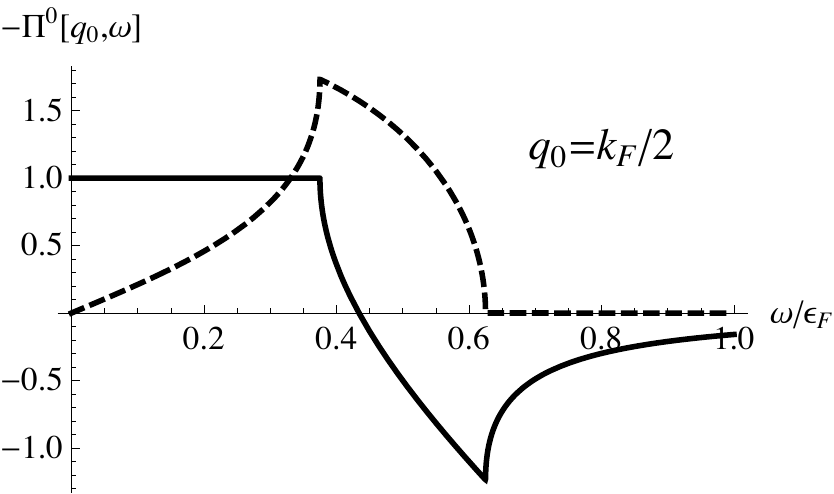}}
  \subfigure[]{\label{Pi0B02DEG-kF2_5}\includegraphics[width=0.23\textwidth]{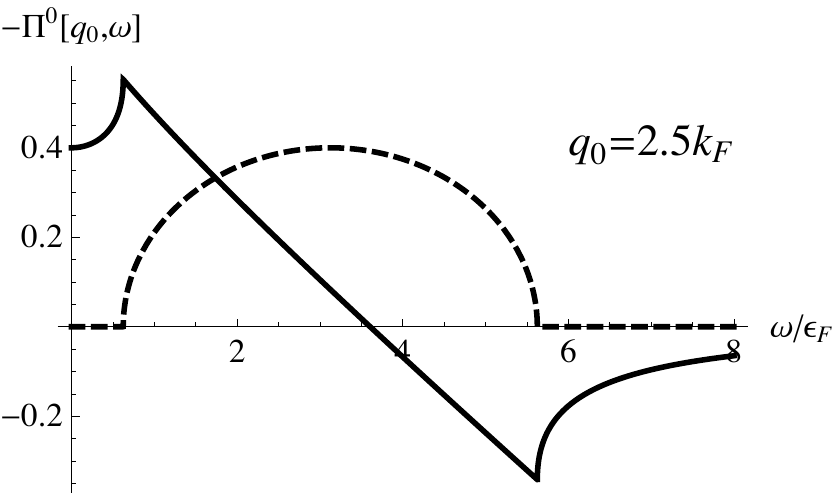}}
   \subfigure[]{\label{Pi0B0-kF0_5}\includegraphics[width=0.24\textwidth]{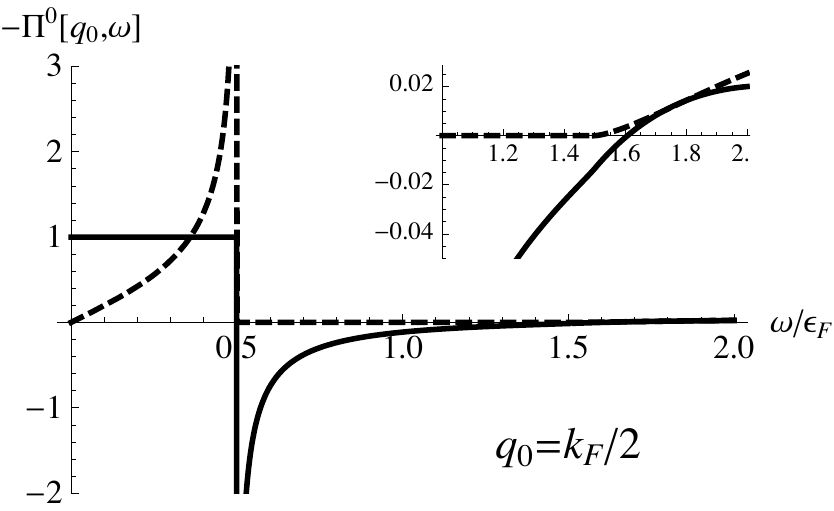}}
  \subfigure[]{\label{Pi0B0-kF2_5}\includegraphics[width=0.23\textwidth]{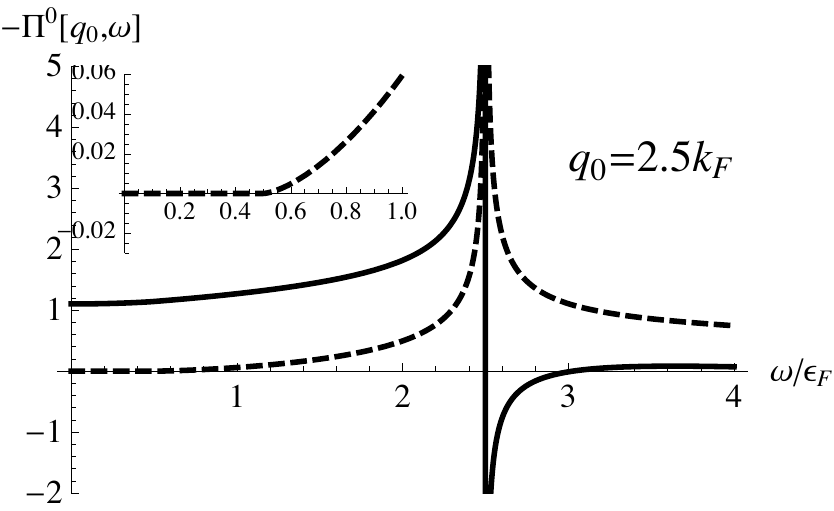}}
  \caption{Polarization function of a 2DEG [plots (a) and (b)] and graphene [(c) and (d)] for different wave-vectors $q_0$. Solid lines set for ${\rm Re}\,\Pi^0(q_0,\omega)$ and dashed lines for ${\rm Im}\,\Pi^0(q_0,\omega)$. $q_0=0.5k_F$ in (a) and (c), and $q_0=2.5k_F$ in (b) and (d). All the plots are done for $\delta\rightarrow 0^+$.}
  \label{Fig:Pi0(q)}
\end{figure}

The real and imaginary parts of the polarization function are
related via the Kramers-Kronig relations. Their behavior is shown in
Fig. \ref{Fig:Pi0(q)} where we plot $-\Pi^0(q_0,\omega)$ for
different wave-vectors $q_0$ for both, 2DEG [plots (a) and (b)] and
graphene [plots (c) and (d)]. We note that most of the spectral
weight (region with non-zero ${\rm Im}\,\Pi^0$) of graphene is
concentrated near $\omega = \vf q$ (due to the absence of
backscattering), whereas in the 2DEG it is more uniformly
distributed.

When electron-electron interactions are included in the problem, RPA
has been shown to capture the essential physics for doped
graphene:\cite{WSSG06,HS07,SNC08} contrary to undoped graphene, the
density of states is finite at the Fermi level once doping moves
$\ef$ away from the Dirac points. However, due to the vanishing density of states at the Dirac point,\cite{K06} the RPA description is
incomplete for undoped graphene,\cite{GFM08} and it is therefore
necessary to consider another class of diagrams in perturbation
theory (such as ladder-type vertex corrections) which lead to a plasmon
resonance below the threshold $\omega<\vf q$. For doped graphene,
the poles of $\Pi^{RPA}(\bq,\omega)$ define the dispersion of a
collective plasmon mode, which has the same long-wavelength
$\sqrt{q}$ behavior as the plasmon in the 2D electron gas studied in
the preceding section [see Eq. (\ref{EqPlasmonB02DEG}) and Fig.
\ref{ImPiRPAB0DP}]. An approximate dispersion relation of the
plasmon in a single layer of doped graphene was calculated in Ref.
\onlinecite{S86} in the framework of a study of intercalated
graphite and was found to coincide with Eq. \ref{EqPlasmonB02DEG},
in which the Fermi energy is now related to the uniform density of
electrons $n_{el}$ as $\ef=\vf k_F=\vf \sqrt{\pi n_{el}}$. Within
the previous approximation, the plasmon mode enters the inter-band
region of the PHES at a momentum
\begin{equation}\label{Eq:qcgraph}
q_c=4k_F \left (2+r_s-\sqrt{r_s^2+4r_s+3}\right),
\end{equation}
where $r_s\equiv e^2/\varepsilon_b\vf$. Note that $r_s$ is density-dependent in the 2DEG whereas it is scale-invariant in graphene. One
important difference with respect to the 2DEG case is that only for
$\omega>\vf q$ it is possible to have a solution for the plasmon
dispersion $1=v(q){\rm Re}\,\Pi^0(q,\omega)$, because ${\rm
Re}\,\Pi^0(\bq,\omega)<0$ for $\omega<\vf q$ [Fig.
\ref{Fig:Pi0(q)}(c)-(d)]. As a consequence, the collective plasmon
mode in graphene at zero magnetic field can only be damped when
decaying into inter-band (and never into intra-band) particle-hole
excitations. Notice also the difference with respect to the 2DEG
case, where the plasmon mode is damped once its dispersion touches,
at some critical wave-vector, the boundary of the PHES. The RPA
dispersion relation of the 2DEG plasmon itself, however, never
enters the intra-band continuum. In the case of graphene, even if
the plasmon enters the inter-band region of the PHES at a well
defined wave-vector approximately given by Eq. (\ref{Eq:qcgraph}),
the mode continues to exist in a rather large region of the
inter-band continuum.\cite{PM08} Recently it has been
argued\cite{PMV09} that, due to the lack of Galilean invariance in
graphene, exchange interactions, which are not included in the RPA,
renormalize the plasmon dispersion of doped graphene in the long-wavelength limit. This renormalization is due primarily to non-local
inter-band exchange interactions, which reduce the plasmon frequency
relative to the RPA value.

We study now the PHES of doped graphene in the presence of a strong magnetic field perpendicular to the sample.
As in the zero magnetic field case, the strong
contribution to the polarization comes from the divergence of
$\Pi^0(\bq,\omega)$ at $\omega=\vf q$, as it can be seen in
Fig. \ref{ImPi0DP}.  The effect of the LL wave-function overlap ${\cal\overline{F}}_{nn^{\prime}}^{\lambda\lambda^{\prime}}(\bq)$ [Eq. (\ref{FF})] is appreciable, in the intra-band region of the spectrum in Fig. \ref{PHESzoom}(b), where we see that the intensity of the modes is larger near the threshold $\vf q$ and practically unappreciable near the second boundary of the intra-band PHES $\omega=\max(0,\vf q-2\ef)$. But in addition, ${\rm Im}\,\Pi^0(\bq,\omega)$
is finite not only in the intra-band region, but also in the zones of Fig. \ref{ImPi0DP} with a finite weight,
as the yellow stripes above and below $\omega=\vf q$. This form of the PHES is due to both, the LL
structure of the spectrum (\ref{relLL}) and the presence of inter-band
excitations that lead to the mentioned stripes in region II of the PHES. The most salient feature that we find comparing the PHES of a 2DEG [Fig. \ref{ImPi0DP2DEG}] to that of graphene [Fig. \ref{ImPi0DP}] in a magnetic field, is that in the former the spectrum is composed of horizontal and equidistant non-dispersive lines, while in the latter these modes are not visible, and the important modes are the diagonal lines parallel to the threshold $\omega=\vf q$.

This particular feature of the PHES of graphene in a strong magnetic
field may be understood in the following manner. Notice first that,
in contrast to the 2DEG with its equally spaced LLs, the spacing of
the relativistic LLs in graphene (\ref{relLL}) decreases at higher
energies. In a fixed energy window at high energies, there are
therefore more possible inter-LL excitations from the level $n$ in
the band $\lambda$ to $n'$ in the conduction band, of energy
$\omega_{n,n'-n}^{\lambda}=\sqrt{2}(\vf/l_B)(\sqrt{n'}-\lambda\sqrt{n})$,
than at lower energies. Notice further that above an energy of
 $\epsilon_F$ also inter-band transitions with $\lambda=-$
contribute. Even for small values of $\delta$, i.e. in clean
samples, neighboring LL transitions overlap in energy such that the
horizontal lines, which dominate the PHES of the 2DEG in a strong
magnetic field [Fig. \ref{ImPi0DP2DEG}], are blurred. At fixed
energy, the spectral weight is again not homogeneously distributed.
This is a consequence of the wave-function overlap between the
electron and the hole, as we discuss in more detail in the following
section.

The above-mentioned dispersive modes in graphene, acquire coherence
once electron-electron interactions are taken into account. This can
be seen in Fig. \ref{ImPiRPADP}, which shows the RPA polarization
function, and where the dispersive diagonal lines are now clearly
distinguishable. We will refer to them as {\sl linear
magneto-plasmons}. In the inter-band region of the PHES, the number
of linear magneto-plasmons depends on the high energy cutoff $N_c$.
We should keep in mind that the RPA is a good approximation for
describing the long-wavelength part of the spectrum, but fails in
reproducing many of the physical properties of a system in the
short-wavelength regime. The dispersion of the collective modes at short-wavelength is renormalized when
diagrammatic contributions beyond the RPA are taken into account. In
particular, the inclusion of ladder diagrams, which account for the
direct interaction between the electron and hole, as well as the
exchange terms, lead to the excitonic and exchange shifts in the
magneto-exciton dispersion.\cite{KH84,IWFB07,BM08}

\section{Structure of the particle-hole excitation spectrum: a wave-function analysis}\label{Semiclassic}

\begin{figure}[t]
  \centering
  \subfigure[]{\label{ImPi0B02DEGIslands}\includegraphics[width=0.24\textwidth]{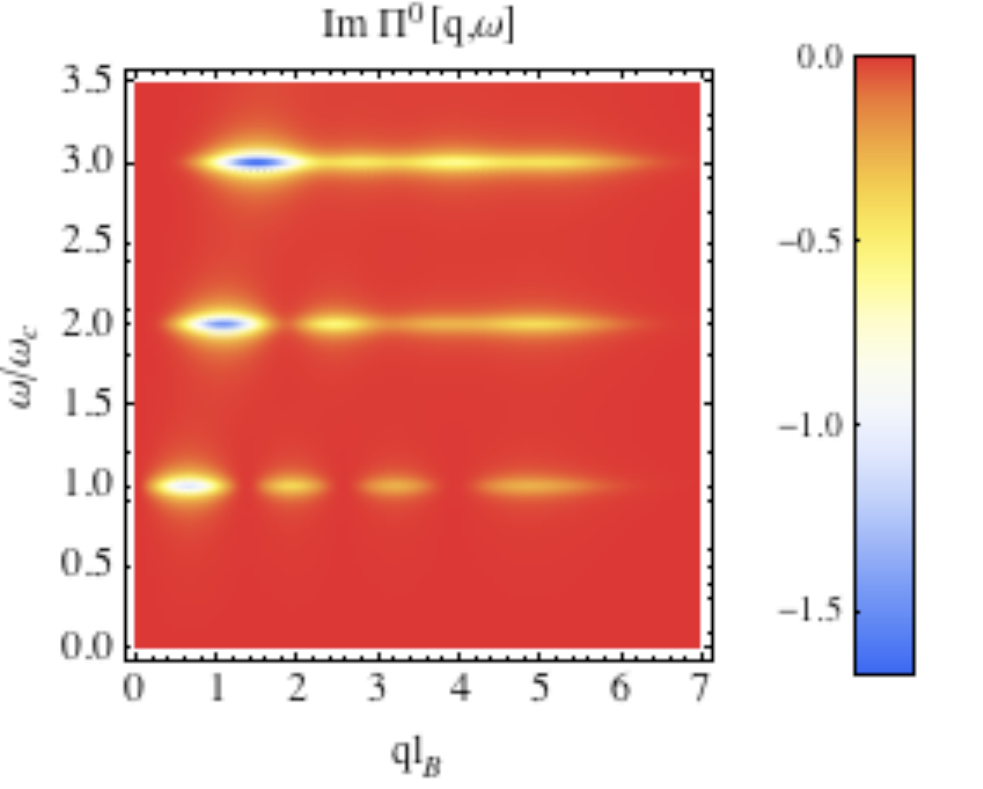}}
  \subfigure[]{\label{ImPi0zoom_delt0_05}\includegraphics[width=0.23\textwidth]{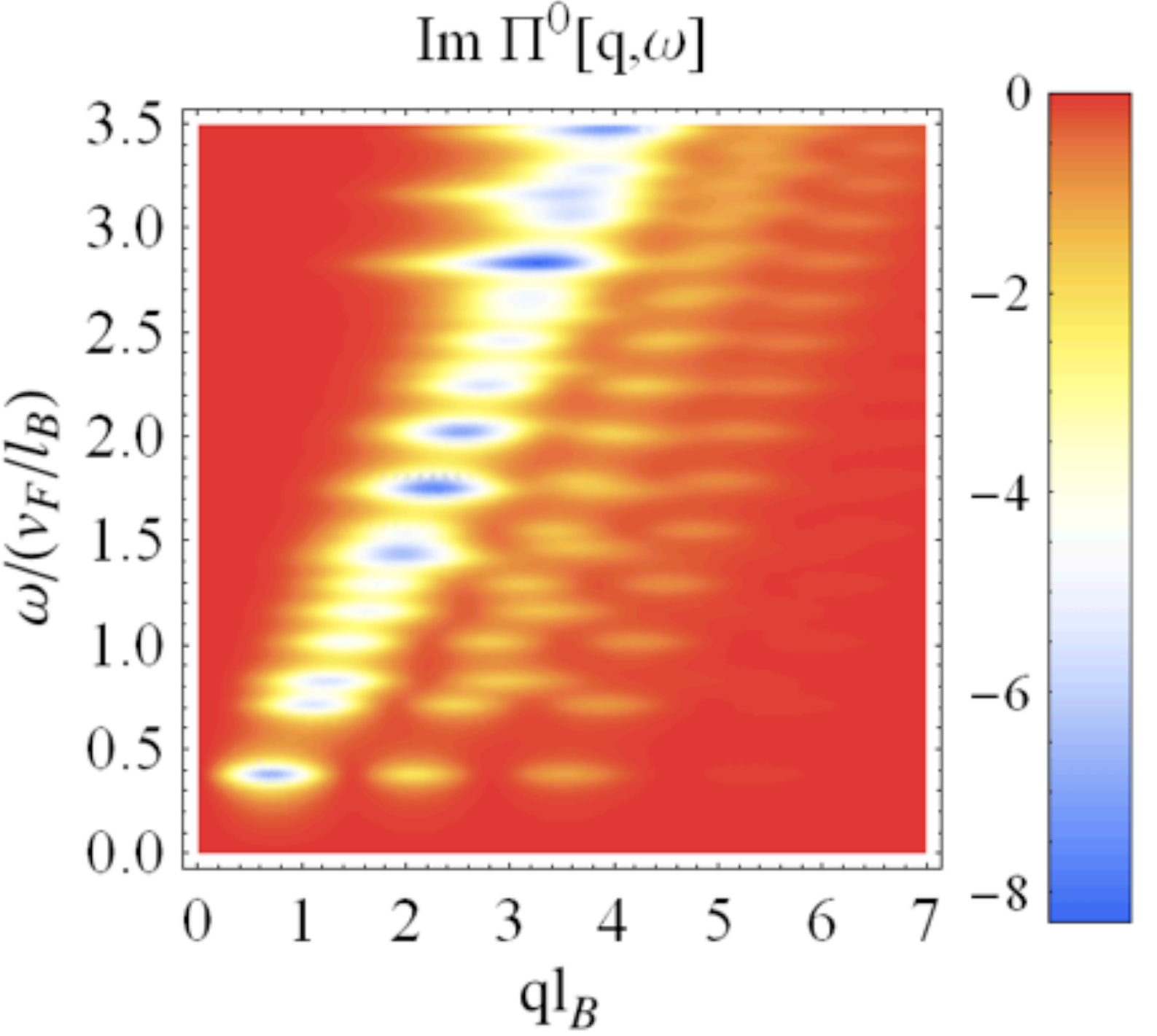}}
  \caption{Density plot of ${\rm Im}\,\Pi^{0}(\bq,\omega)$ in the low energy sector.
  Plot (a) corresponds to a 2DEG and plot (b) to graphene. We have used $N_F=3$ and $\delta=0.05$ in units of $\omega_c$ and $\vf/l_B$, respectively. {\it Islands} (see text) are clearly visible in the $\omega=\omega_c$ horizontal line in (a) and in every horizontal line in (b).}
  \label{PHESzoom}
\end{figure}

The boundaries of the PHES in a magnetic field may be understood by
considering an electron-hole pair and treating the cyclotron motion
of both the electron and the hole in a semiclassical limit. The
boundaries are related to the region in real space where the
electron and hole cyclotron orbits may overlap. If we decompose the
position of an electron $\br$ into its cyclotron $\pmb \eta$ and
guiding center $\bf R$ coordinates, $\br={\bf R}+{\pmb \eta}$, the
finite overlap of the electron and hole orbits implies that
$\eta_c'-\eta_c\le \Delta R \le \eta_c'+\eta_c$, where $\Delta {\bf
R}={\bf R}'-{\bf R}$, $\Delta R=|\Delta {\bf R}|$ and $\eta_c'$ and
$\eta_c$ are the cyclotron radius of the electron and the hole,
respectively. The latter are given by $\eta_c\equiv \langle | {\pmb
\eta}|\rangle=l_B\sqrt{2n+1}$ and $\eta_c' =l_B\sqrt{2n'+1}$ in
terms of the LL indices $n$ and $n'$ of respectively the hole and
the electron. As the distance between guiding centers $\Delta R$ is
related to the electron-hole pair momentum ${\bf q}$ by $\Delta R=q
l_B^2$ (see Appendix \ref{App:EHPM}), the momentum is constrained to
\begin{equation}\label{EqBoundaries}
\sqrt{2n'+1}-\sqrt{2n+1}\le ql_B\le \sqrt{2n'+1}+\sqrt{2n+1}.
\end{equation}
As an illustration, the boundaries of the PHES at $m\equiv n'-n=1$
and for $N_F=3$ obtained from Eq. (\ref{EqBoundaries}), $0.35\le q
l_B\le 5.65$, coincide with that shown in Fig. \ref{PHESzoom}(a) and
(b). 

\begin{figure}[t]
  \centering
  \subfigure[]{\label{ImPi02DEG_omega1}\includegraphics[width=0.23\textwidth]{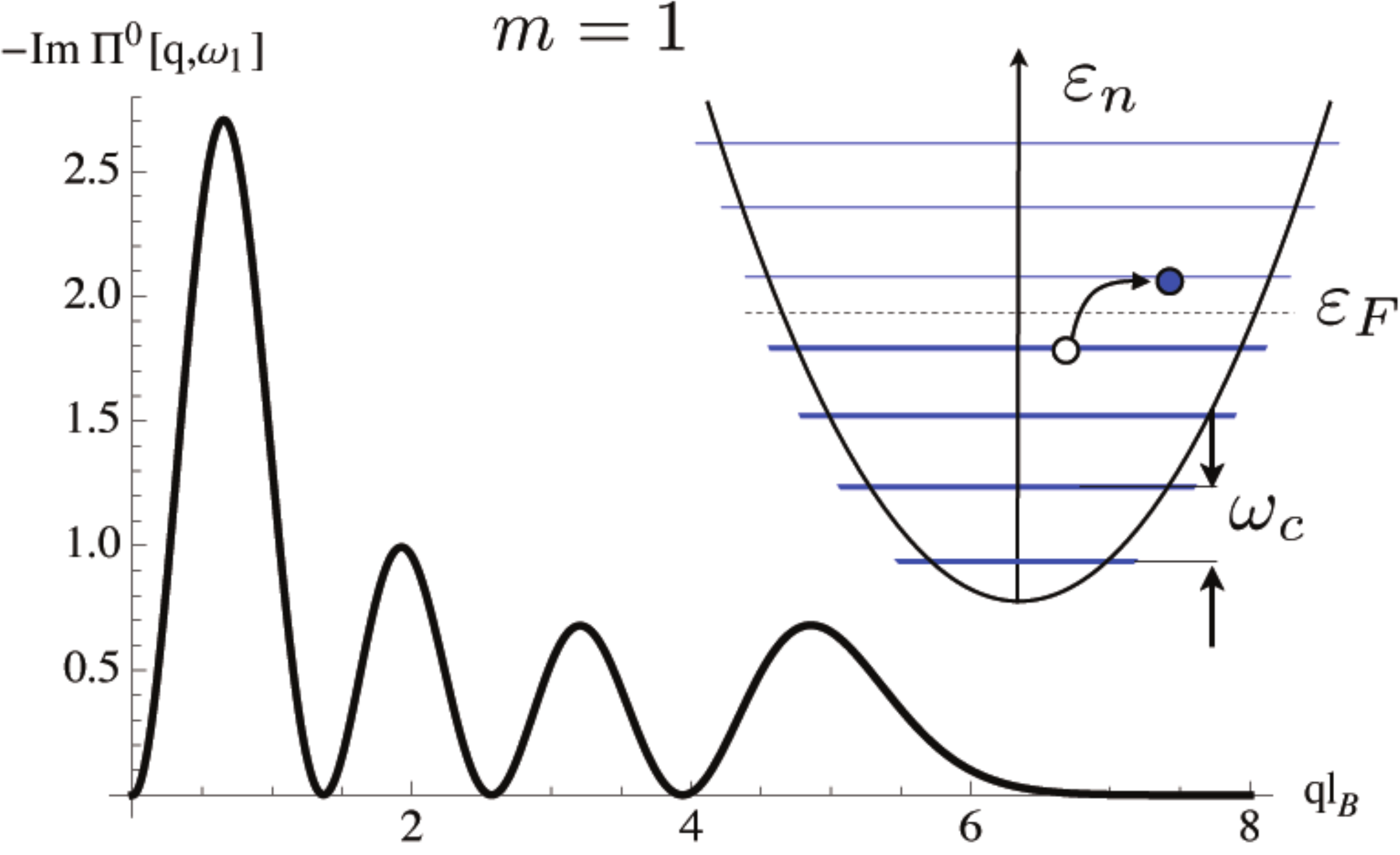}}
  \subfigure[]{\label{ImPi0_omega1}\includegraphics[width=0.23\textwidth]{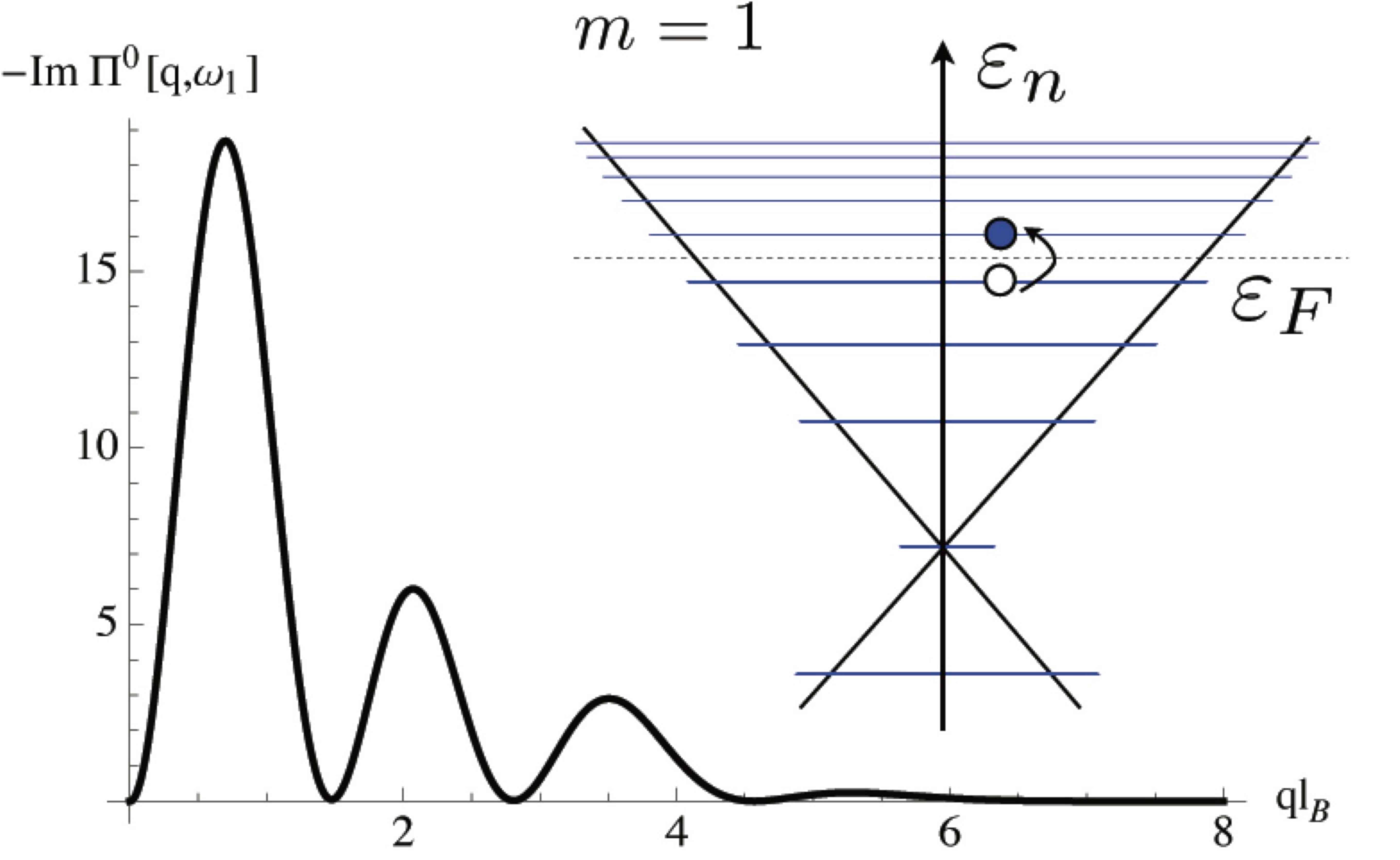}}
  \subfigure[]{\label{ImPi02DEG_omega2}\includegraphics[width=0.23\textwidth]{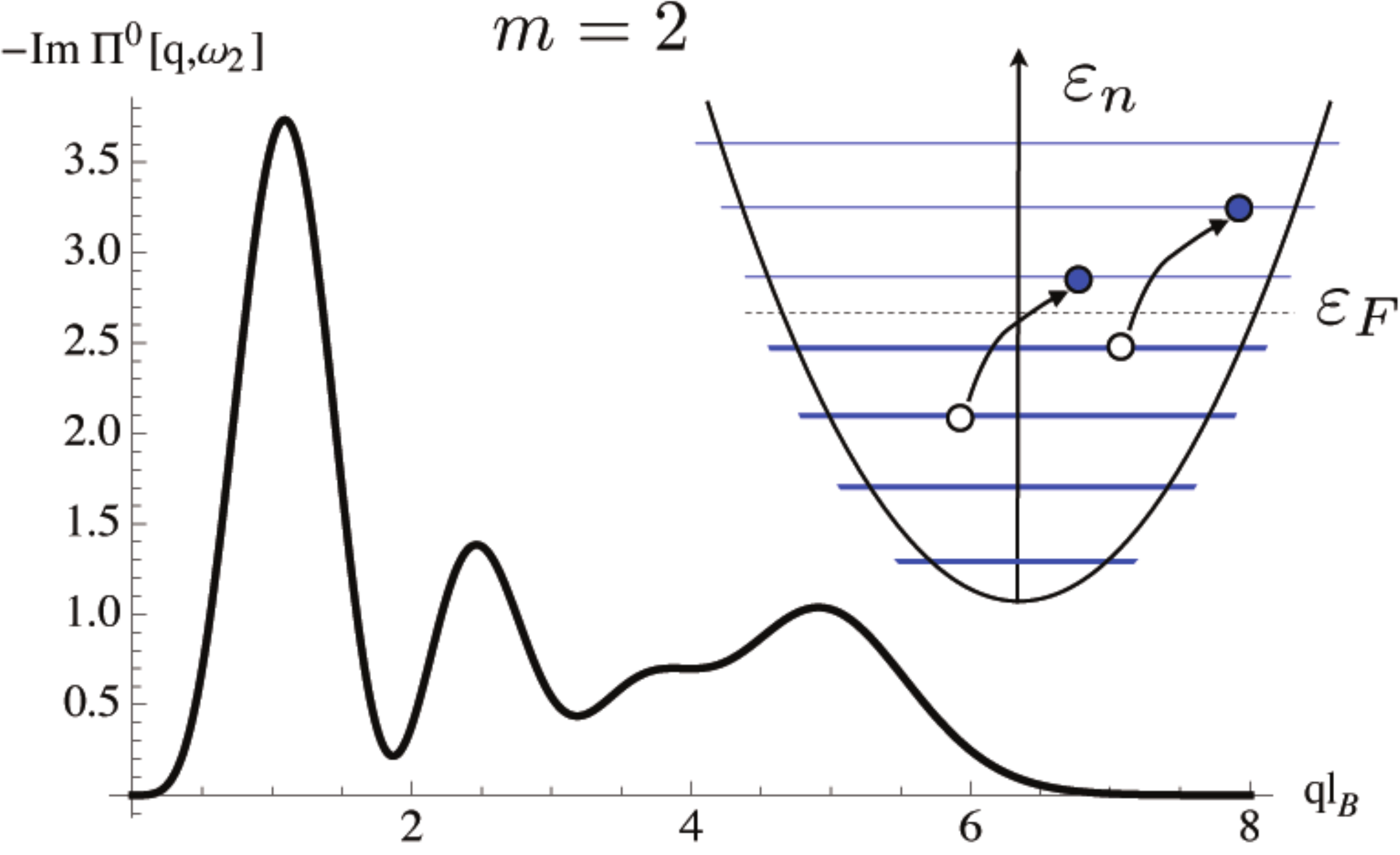}}
  \subfigure[]{\label{ImPi0_omega2}\includegraphics[width=0.23\textwidth]{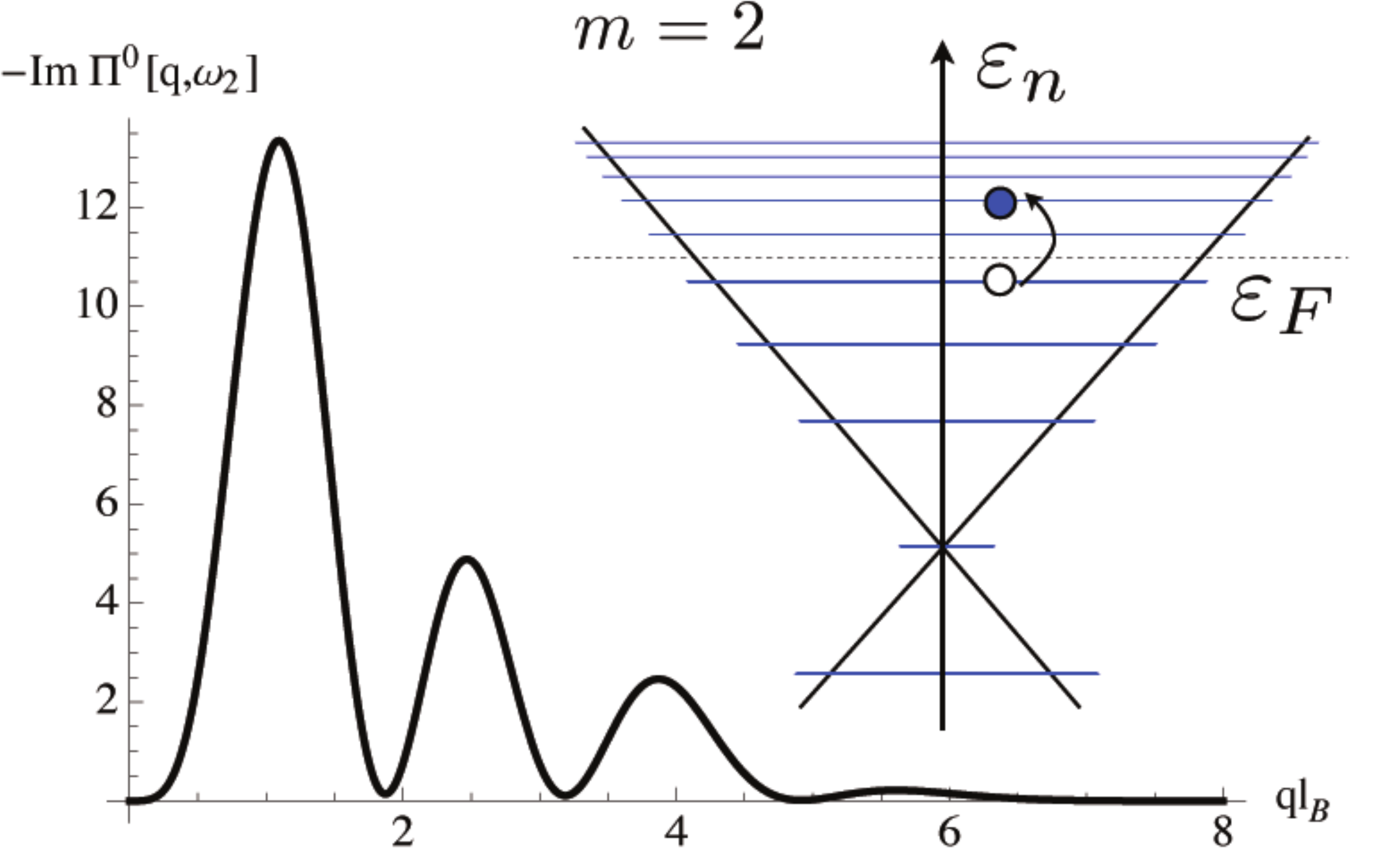}}
    \subfigure[]{\label{ImPi02DEG_omega3}\includegraphics[width=0.23\textwidth]{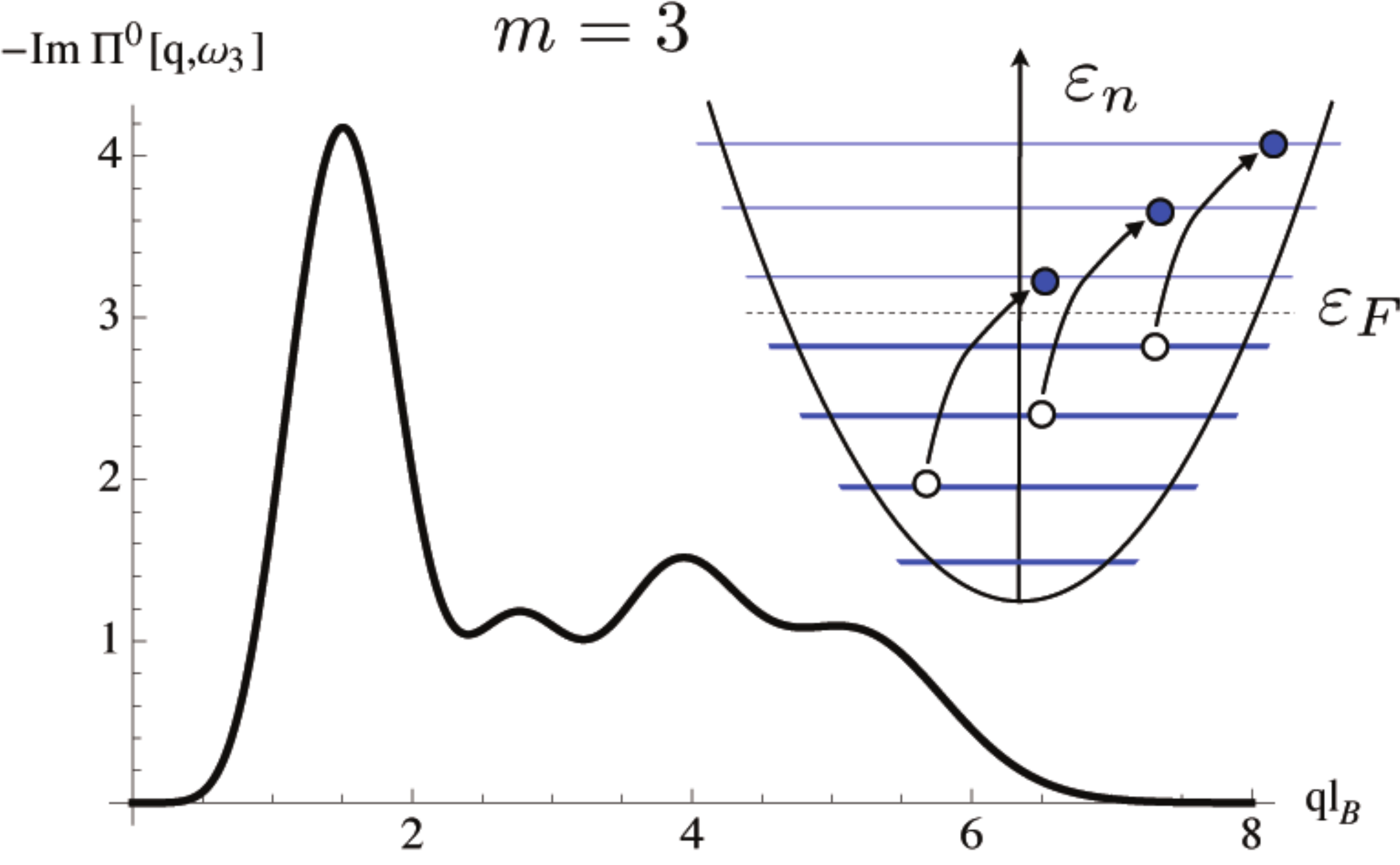}}
  \subfigure[]{\label{ImPi0_omega3}\includegraphics[width=0.23\textwidth]{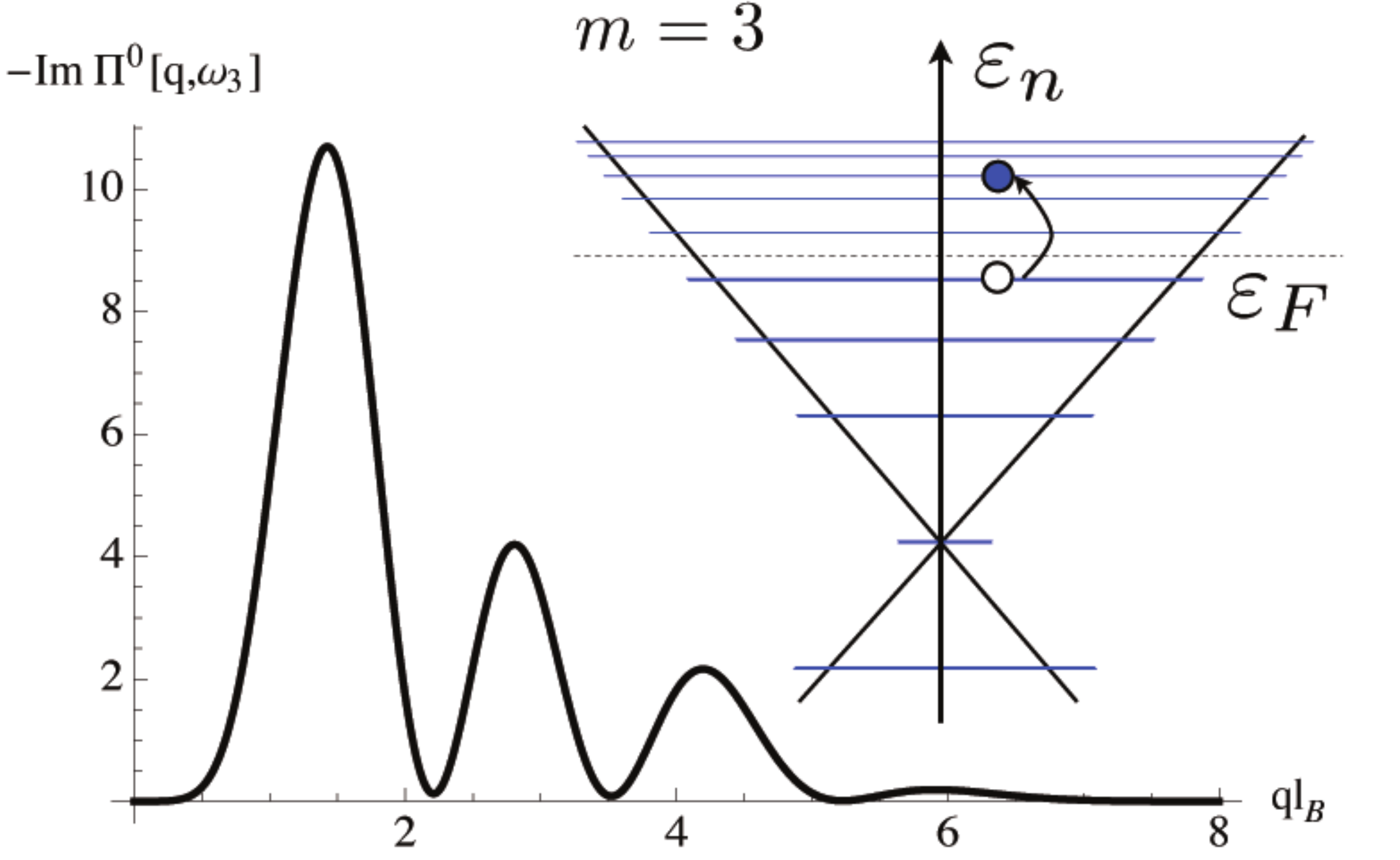}}
  \caption{${\rm Im}\,\Pi^{0}(\bq,\omega_m)$ for a 2DEG [plots (a), (c) and (e)] and
  for graphene [plots (b), (d) and (f)]. Here $\omega^{2DEG}_m=m\,\omega_c$ in a 2DEG and
  $\omega^{graph}_m=\vf l_B^{-1}(\sqrt{2(N_F+m)}-\sqrt{2N_F})$ in graphene. The
  inset of each figure represents the different particle-hole excitations contributing at
  that energy. As we use $N_F=3$ for all the plots, there are $N_F+1=4$
  maxima in ${\rm Im}\,\Pi^{0}(\bq,\omega_m)$ corresponding to
  {\it islands} (see text).
  Notice that for graphene, we show the
  polarizability at the energy of the first, second and fourth horizontal line of Fig.
   \ref{ImPi0zoom_delt0_05}. The third line at this filling corresponds to the
   $(N_F+1,N_F-1)$ electron-hole pair and presents therefore $N_F$ peaks.}
  \label{ImPi0_omegan}
\end{figure}

The presence of a set of {\it islands} -- which is the name we give
to regions of high spectral weight within each electron-hole contribution $(n+m,n)$
to a given horizontal line $m$ -- can be understood
from the form of the wave functions of the electron and the hole forming the pair. In the symmetric gauge, the
modulus $|\Psi_{n,\ell}(x,y)|$ of the LL wave-function is rotation-invariant, its shape
being that of $n+1$ concentric and equidistant rings (of average
radius $r_{\ell}\simeq \sqrt{2\ell}l_B$).\cite{GV05} Therefore, one expects
that electron-hole excitations of momentum $q$ will be possible
whenever there is a finite overlap of the particle and hole
wave-functions, the guiding centers of which are separated by a
distance $\Delta R=l_B^2q$. If $n'$ is the LL index of the electron
and $n$ that of the hole, there are $n+1$ substantial overlaps of the rings of
the electron and the hole wave functions when $\eta_c'-\eta_c\le
\Delta R \le \eta_c'+\eta_c$. This will lead to a division of the
contribution $(n',n)$ to the $m$-th horizontal line of the PHES into
$n+1$ regions of preferred momenta or islands.

In order to understand the effect of these overlap functions in more
detail, we first consider the line $m=1$ of lowest energy. Both in
the 2DEG and in graphene, the only possible inter-LL excitation that
contributes to the formation of this energy line involves a hole in
the LL $N_F$ and an electron in $N_F+1$. This is schematically
represented in the inset of Fig. \ref{ImPi02DEG_omega1}-(b), where
we represent the unique electron-hole transition contributing to the
first horizontal line in each case.  As argued above, there are
$N_F+1$ zones of preferred momenta due to the overlap between the
wave functions of the electron and the hole. In Fig.
\ref{ImPi0_omegan}(a) and (b), we have plotted ${\rm
Im}\,\Pi^{0}(\bq,\omega_m)$ for the 2DEG and graphene, respectively,
at the energy corresponding to the first horizontal line, with
$N_F=3$. One obtains indeed four peaks which yield the islands
observed in the low-energy zoom of the PHES (see Fig. \ref{PHESzoom}
where we have also chosen $N_F=3$). Notice, however, that the last
island, though present, is hardly visible in graphene due to the
suppression of backscattering.

At larger values of $m$, there is an essential difference between the 2DEG and graphene. Due to the equidistant
level spacing in LL quantization for non-relativistic electrons in the 2DEG, the horizontal lines in the PHES occur at
$\omega_m=m\omega_c$. If $m>1$, there are $\min(m,N_F)$ different inter-LL transitions $(n+m,n)$ that contribute to the spectral weight
of the $m$-th horizontal line [see the inset of Fig. \ref{ImPi02DEG_omega2} and (e) which indicates the electron-hole transitions contributing to the second and third horizontal line in a 2DEG]. But all these transitions have different preferred momenta because of different electron-hole
overlaps. The spectral weight is therefore a superposition of these different overlap functions [see  Fig. \ref{ImPi0_omegan}(c), (e),
where we plot ${\rm Im}\,\Pi^{0}(\bq,\omega_m)$ for $m=2,\,3$] and the islands are no longer well defined in the horizontal
direction, as it may be seen in the second and third horizontal line of Fig. \ref{ImPi0B02DEGIslands}, which have lost the dashed structure of the first line. The situation is remarkably different in graphene, where the LL spacing is not constant and where
a particular horizontal line is due to the inter-LL
transition with energy $\omega_{n,m}^{\lambda}=\sqrt{2}(\vf/l_B)(\sqrt{n+m}-\lambda\sqrt{n})$ and therefore not only determined
by the LL-index separation $m$. 
Apart from very rare events in the high-energy regime where two
inter-LL transitions $(n_1+m_1,n_1)$ and $(n_2+m_2,n_2)$ may
coincide in energy
$\omega_{n_1,m_1}^{\lambda}=\omega_{n_2,m_2}^{\lambda'}$, each
horizontal line therefore consists of a {\sl single} inter-LL
transition and has $N_F+1$ well-separated peaks in ${\rm
Im}\,\Pi^{0}(\bq,\omega_{n,m}^{\lambda})$, as we have shown in Fig.
\ref{ImPi0_omegan}(d), (f) for $(N_F+m,m)$, with $m=2,3$. In contrast to
the 2DEG, the islands remain thus well separated in the horizontal
direction, whereas they overlap strongly in the vertical direction
(i.e. in energy) due to the decreasing level spacing at higher
energies and the large number of inter-LL transitions in a fixed
energy window [Fig. \ref{ImPi0zoom_delt0_05}], as we have discussed in the last section. As a
consequence, the most prominent modes in graphene in a strong
magnetic field are diagonal lines, parallel to $\omega=\vf
q$, whereas those in the 2DEG remain horizontal. Electron-electron
interactions turn these lines of large spectral weight into {\sl
coherent} modes: magneto-excitons in the 2DEG and linear
magneto-plasmons in the case of graphene, in addition to the upper
hybrid mode that reveals itself in the formerly forbidden energy
region of the PHES of non-interacting particles.

\section{Static screening}

\begin{figure}[h]
\centering
  \subfigure[]{\label{Pi0DescB0}\includegraphics[width=0.24\textwidth]{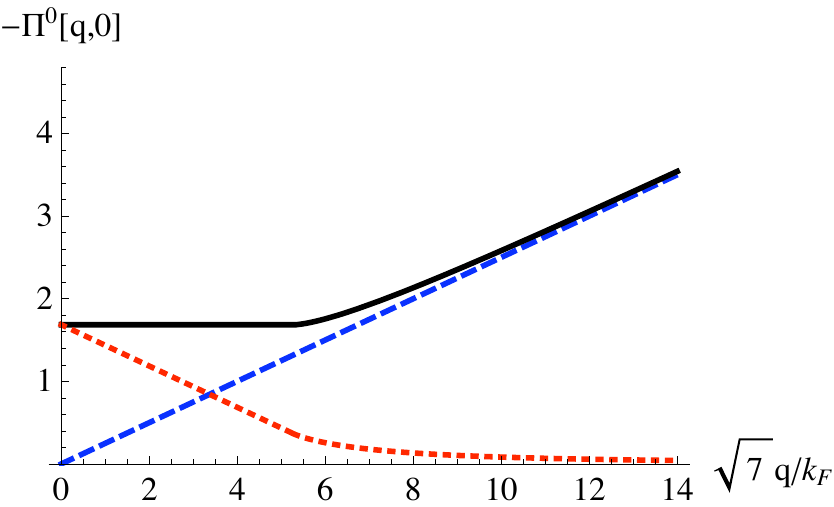}}
  \subfigure[]{\label{Pi0Desc}\includegraphics[width=0.23\textwidth]{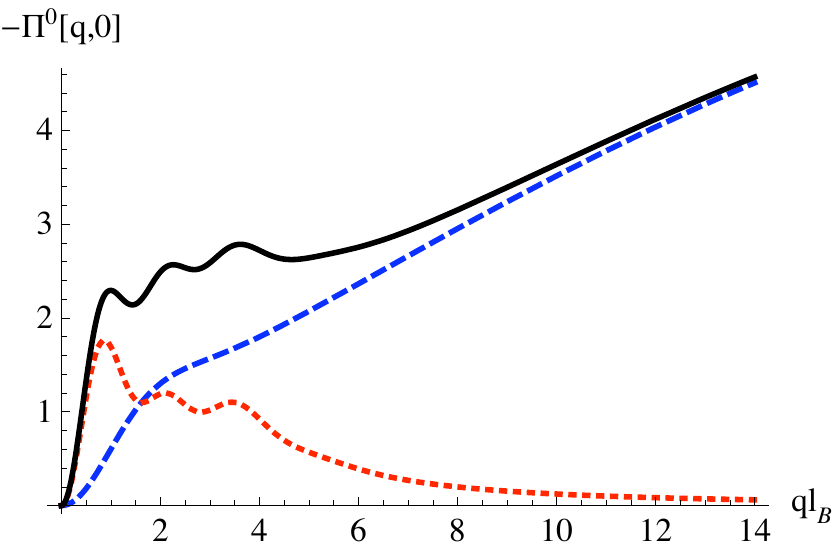}}
   \subfigure[]{\label{Pi0B02DEG}\includegraphics[width=0.24\textwidth]{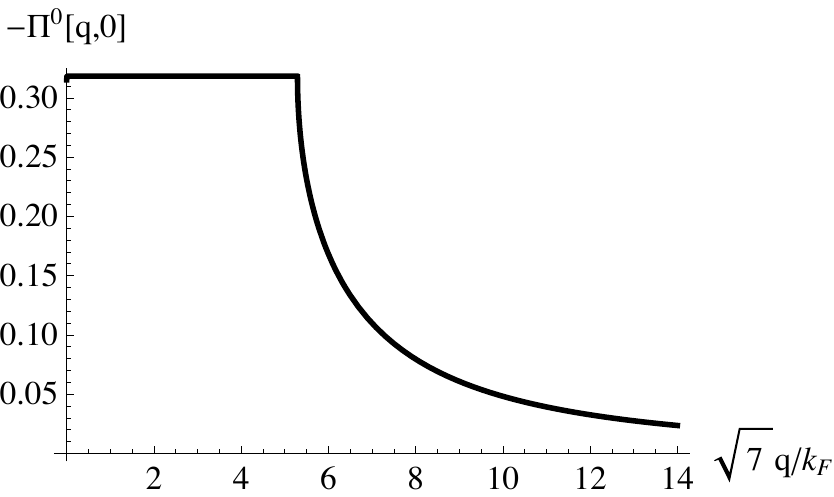}}
  \subfigure[]{\label{Pi02DEG}\includegraphics[width=0.23\textwidth]{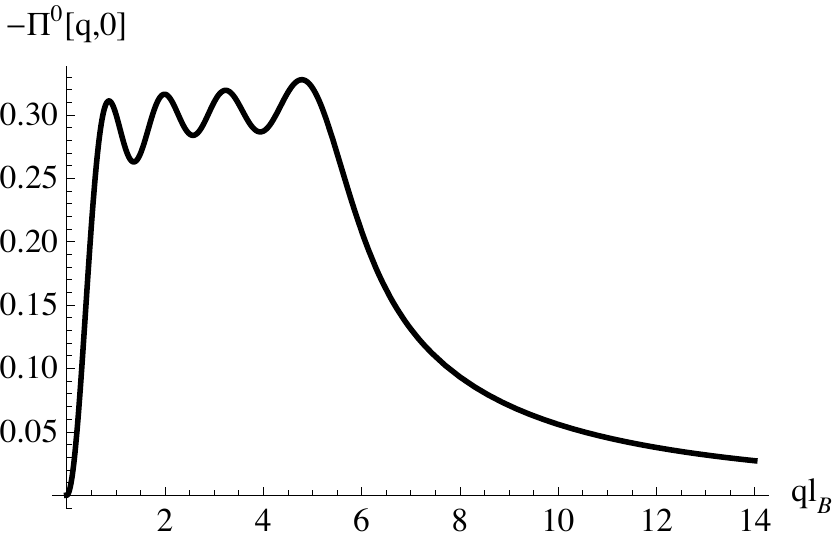}}
\caption{(Color online) (a) Vacuum (inter-band) contribution (blue dashed line), intra-band
contribution (red dotted line) and total static polarization
function (black thick line) of graphene for $B=0$ (see Ref.\onlinecite{HS07}). (b)
Same as (a) but for $B\ne 0$. We have used $N_F=3$ and $N_c=350$.
For consistency, the Fermi momentum used in (a) is
$k_F=\sqrt{2N_F+1}/l_B=\sqrt{7}/l_B$. (c)-(d) Same as (a)-(b) respectively, but for a standard 2DEG.}\label{PiStatic}
\end{figure}

In this section we study the properties of $\Pi^0(\bq)=\Pi^0(\bq,\omega=0)$
in the static limit, for which the polarization is entirely real.
The polarizability of graphene is shown in Fig. \ref{PiStatic}(a)
and (b) for $B=0$\cite{S86,WSSG06,HS07}  and $B\neq 0$,
respectively. For comparison, we also show the corresponding
polarizability of a standard 2DEG \cite{GV05} in the absence [Figs.
\ref{Pi0B02DEG}] and in the presence [Fig. \ref{Pi02DEG}] of a
magnetic field. In order to compare the $B\neq 0$ to the $B=0$
polarizability, we have chosen a Fermi wave-vector $k_F$ for the
$B=0$ case equal to $\sqrt{2N_F+1}/l_B$, which corresponds to the
same carrier density as a graphene layer in a magnetic field with
all LLs filled up to the $N_F$-th level of the
conduction band.

One first notices a difference in the $B=0$ static polarizabilities
between graphene and the 2DEG. Although the static polarizability
remains constant and equal to the electronic density of states
$\rho(\epsilon_F)$,\footnote{The density of states (per unit area) at the Fermi
energy is a constant equal to $gm_b/2\pi$ for a 2DEG, where
$g=g_s=2$ accounts for the spin degeneracy, whereas for graphene it
is energy dependent and given by $gk_F/(2\pi \vf)$, where
$g=g_sg_v=4$ accounts for spin and valley degeneracy.} in both
cases up to a wave vector $2k_F$, there are two contributions for
graphene that stem from intra-band and inter-band excitations,
respectively. Whereas the polarizability due to intra-band
excitations in graphene [red dotted line in Fig. \ref{PiStatic}(a)]
decreases linearly in $q$, due to the electrons' chirality (\ref{OverlapB0}) and the absence of backscattering, the
intra-band contributions yield a linearly increasing polarizability.
Beyond $q=2k_F$, there are no possible zero-energy particle-hole
excitations in the intra-band region, and the associated
polarizability therefore tends to zero. This is also the case in the
2DEG [Fig. \ref{Pi0B02DEG}], where there are only intra-band
excitations. In graphene, however, inter-band excitations still
yield a linearly increasing contribution to the total
polarizability, which then asymptotically approaches the inter-band
polarizability [blue dashed line in Fig. \ref{PiStatic}(a)].

Qualitatively, one finds a similar behavior for the static
polarizability at $B\neq 0$ except in the
small-$q$ limit. Whereas the static polarizability at $B=0$ remains
constant and coincides with the density of states at the Fermi
energy, it tends to zero as $\Pi^0(\bq\rightarrow 0)\propto q^2$ for
$B\neq 0$.\cite{GV05} This is due to the fact that the main contribution to
the polarizability comes from $\bq=0$ excitations in the vicinity of
the Fermi energy $\epsilon_F$. Contrary to the $B=0$ case, where
there are $\bq=0$ excitations the energy of which tends to zero,
$\epsilon_F$ lies now in the cyclotron gap between the highest occupied LL
$N_F$ and the lowest unoccupied one $N_F+1$. This energy gap must be
overcome by $\bq=0$ excitations, such that its spectral weight tends
to zero then. Indeed, the static polarizability also coincides with
the density of states at the Fermi energy because the latter
vanishes for $B\neq 0$ when $\epsilon_F$ lies in the gap.

Furthermore, one notices the oscillatory behavior of the static polarizability, both for graphene and the 2DEG, below
$2k_F$. These oscillations are again due to the wave-function overlap between the electron and the hole, and one
obtains $N_F+1$ maxima. Since the main contribution to the polarizability at small wave vectors comes from excitations in
the vicinity of $\epsilon_F$, the oscillations are dominated by the $(N_F+1,N_F)$ intra-band transition in graphene,
as one may also see from the red dotted line in Fig. \ref{PiStatic}(b), which represents the intra-band contribution to
the polarizability. At large values of the wave vector $\bq$ the static polarizability is, as in the $B=0$ case, dominated
by inter-band excitations the discrete nature of which is less important than in the small-$\bq$ limit. The linear increase
therefore coincides with the $B=0$ result.

\begin{figure}[h]
  \centering
  \subfigure[]{\label{EpsStatic}\includegraphics[width=0.35\textwidth]{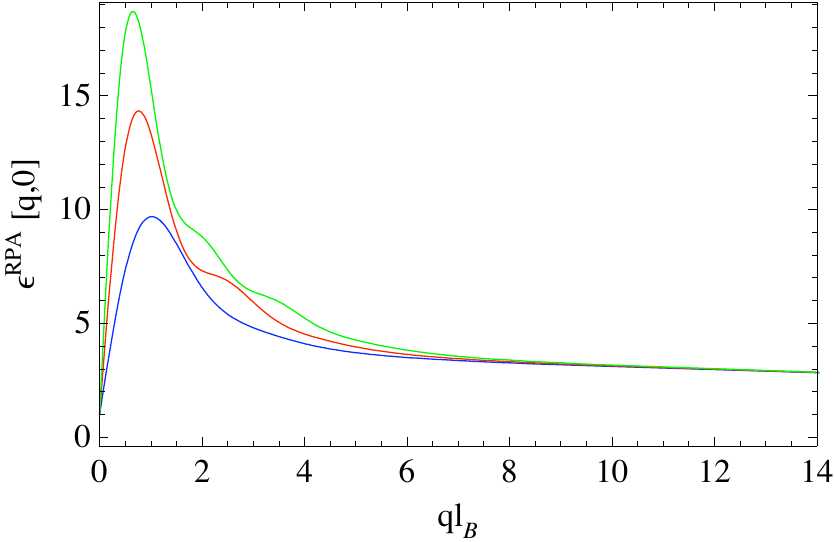}}
   \subfigure[]{\label{Eps2DEGStatic}\includegraphics[width=0.35\textwidth]{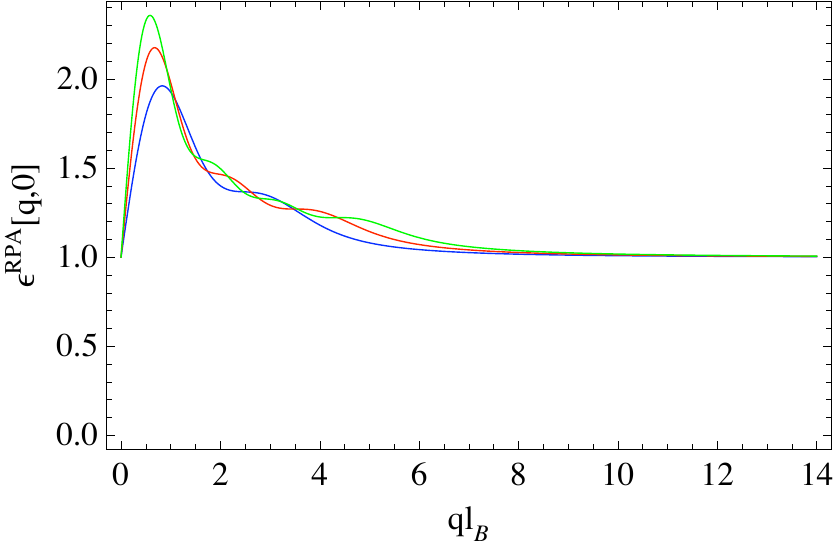}}
  \caption{Static ($\omega=0$) dielectric function for graphene (a) and a 2DEG (b) in a magnetic field computed in the RPA.
  We have used $N_F=1,2,3$, corresponding to the blue, red and green lines, respectively.}
  \label{StaticResults}
\end{figure}

The static polarizability is a useful quantity for the calculation
of the screening properties of electrons. Screening, e.g. of the
Coulomb interaction potential between the electrons or the potential
of a charged impurity, is indeed determined by the (static)
dielectric function $\varepsilon^{RPA}(q,\omega=0)=1-v(q)\Pi^0(q)$.
At zero field, the long wavelength limit is similar in the two
cases: $\varepsilon^{RPA}(q)\approx 1+q_{TF}/q$, where $q_{TF}\equiv
2\pi e^2 \rho(\epsilon_F)/\varepsilon_b$ is the Thomas-Fermi
wave-vector. Note however that $q_{TF}$ is density independent in the
2DEG whereas it scales as $k_F$ in graphene. Therefore, the dielectric function diverges as $\varepsilon \sim
q_{TF}/q\rightarrow \infty$ in the two cases when $q\to 0$. However,
the $k_F$ dependence in the numerator of
$\varepsilon^{RPA}(q\rightarrow 0)$ in graphene points out the
absence of screening in undoped graphene $(k_F=0)$ at long
distances. In the short wavelength region $q\gg 2k_F$,
$\varepsilon^{RPA}(q)$ tends to 1 in a 2DEG, whereas for graphene,
$\varepsilon^{RPA}(q)\rightarrow 1+\pi r_s/2$. This extra
contribution of $\pi r_s/2$ to the dielectric function of graphene
at large wave vectors is due to the linear growth of the
polarizability at $q>2k_F$ and is therefore related to virtual
inter-band particle-hole excitations.\cite{A06} In summary, at short
wavelengths, a 2DEG does not screen at all ($\varepsilon \to 1$),
whereas (doped or undoped) graphene screens as a dielectric
($\varepsilon \to 1+\pi r_s/2>1$) thanks to its filled valence band.

The situation is different in the presence of a magnetic field. In
Fig. \ref{StaticResults}(a) and (b) we have plotted the static
dielectric function for graphene and the 2DEG, respectively, in a
magnetic field (see also Ref. \onlinecite{S07,S08}). Notice that at
long wavelengths, $\varepsilon^{RPA}(q)-1\propto q$ in the 2DEG\cite{AG95} as
well as in graphene. In fact, in the limit of $N_F\gg 1$, 
\begin{equation}
\varepsilon^{RPA}(q)-1\propto r_sN_F^{3/2}ql_B
\end{equation} 
as $q\rightarrow 0$, this limit being valid for both, a 2DEG and graphene. The difference stems again in the density dependence of $r_s$ in the two cases: because $r_s\sim N_F^{-1/2}$ in a 2DEG, $\varepsilon^{RPA}$ grows linearly with $N_F$ in this case. However, $r_s$ is density-independent in graphene, leading to a dielectric function proportional to $N_F^{3/2}$.  Furthermore the maximum of the dielectric function behaves as $\varepsilon_{\max}\simeq \varepsilon^{RPA}(q\sim 1/k_Fl_B^2)\sim r_sN_F$. Therefore $\varepsilon_{\max}\propto \sqrt{N_F}$ in a 2DEG whereas $\varepsilon_{\max}\propto N_F$ in graphene. This different behavior is reflected in Fig. \ref{EpsStatic} and (b). As we see, there is a considerable increase of the static
dielectric function of graphene as we increase $N_F$, as compared to
the 2DEG. This is due to the relativistic LL quantization of
graphene, which leads to an increasing of the quantum effects
(virtual inter-level transitions) as the separation between levels
becomes narrower. On the other hand, in both graphene and the 2DEG,
$\varepsilon^{RPA}(q,\omega=0)\rightarrow 1$ as $q\rightarrow 0$,
which implies that there is no screening at long distances (as in
the undoped zero field case, as discussed above). The short
wavelength behavior of the dielectric function in a magnetic field
is, however, similar in both the 2DEG and graphene to their
respective zero field limits. Therefore, the short wavelength decay
of the effective interaction in graphene, due to inter-band
polarization effects, leads to a screening similar to that  of an
insulator, while the intra-band processes leads to a metallic-like
screening.

\section{Conclusions and Outlook}

In conclusion, we have compared in detail the polarizability of doped graphene with and without a strong magnetic field to
that of the 2DEG. In the absence of a magnetic field, the main difference arises from the presence of two different regions
of non-vanishing spectral weight in the PHES ${\rm Im}\,\Pi(\bq,\omega)$ of graphene. These two regions represent contributions
from intra- and inter-band excitations, respectively, whereas in the 2DEG with only one parabolic band, there is only
one region. Furthermore, the chirality of electrons in graphene suppresses backscattering such that the spectral weight is
centered around the main diagonal of the PHES at $\omega= \vf q$, whereas it is more or less homogeneously distributed over
the particle-hole continuum in the 2DEG. Electron-electron interactions yield in both cases a plasmon mode that disperses as
$\sqrt{q}$.

In the presence of a strong magnetic field, the plasmon mode evolves into the upper hybrid mode which is gapped at zero energy.
In the 2DEG, this gap is given by the cyclotron frequency $\omega_c$, whereas in graphene it is $eB\vf^2/\epsilon_F$ and thus depends
on the Fermi energy $\epsilon_F$. The most salient difference between graphene and the 2DEG are the additional modes that occur
in the interacting system in the parts of the PHES that correspond to the zero-field particle-hole continuum.
In the 2DEG, the spectral weight is concentrated along equidistant horizontal lines, due to the
equidistant LL spacing, and the resulting {\sl magneto-excitons} are therefore weakly dispersing. In contrast to these rather well
studied modes in the 2DEG, one finds {\sl linear magneto-plasmons} in graphene that disperse roughly parallel to the central
diagonal $\omega=\vf q$ in the PHES. These modes are a consequence of a different organization of the regions of highest spectral
weight in graphene as compared to the 2DEG. The energy levels in graphene are no longer equally spaced as a consequence of
relativistic LL quantization, and the energies of inter-LL transitions are more densely packed than in the 2DEG, especially at
higher energies. Even a small level broadening due to impurities therefore leads to an overlap in energy of the inter-LL transitions.
Furthermore, spectral weight is highly modulated at a fixed energy due to the wave-function overlap between the electron and the
hole involved in the excitation.

We have finally discussed the static polarizability and the
dielectric function that describe the screening properties of the
system. In contrast to the 2DEG, where the polarizability tends to
zero at wave vectors larger than $2k_F$, it increases linearly in
graphene, due to the increasing relevance of inter-band excitations.
Whereas the small-$q$ behavior of the dielectric constant is similar
in graphene and the 2DEG, a calculation within the RPA indicates
that it tends to a constant different from one in the large-$q$
limit for graphene.

As for an experimental confirmation of the particular high-field collective excitations in graphene discussed above, one may 
first think of magneto-optical experiments. Transmission spectroscopy has indeed revealed the characteristic $\sqrt{Bn}$ behavior
of the graphene LLs in epitaxial \cite{sadowski06} and exfoliated graphene,\cite{jiang07} in agreement with theoretical 
expecations.\cite{gusynin1,gusynin2,abergel} Similarly, Raman spectroscopy has been successfully applied to graphene in the absence
\cite{yan2007,pisana2007} and in the presence of a magnetic field.\cite{faugeras} 
Thus the particular electron-phonon interaction \cite{andoEP,antonioEP}
and the theoretically studied magneto-phonon resonance\cite{ando2007,goerbigEP,kashuba}
could be confirmed. However, these techniques are restricted to zero wave-vector excitations, 
whereas electron-electron interactions and the resulting collective excitations are more prominent at non-zero values of the wave vector.
In order to probe the excitation spectrum at non-zero values of the wave vector, inelastic light scattering may be a promissing technique
that has been successfully used to study collective quantum-Hall excitations in the 2DEG.\cite{EW99,pinczuk}

\acknowledgments We acknowledge financial support from ``Triangle de
la physique'' and ANR under grant number ANR-06-NANO-019-03.


\appendix

\section{Calculation of the polarization function}
\label{App:Pol}

The electronic wave function in graphene
in a magnetic field can be expressed as a four component spinor,

\begin{equation}\label{psi+}
\Psi_+(\br,t)=
e^{i\bK\cdot\br}\left( \begin{array}{c}
 \Phi_{+A}(\br,t)\\
 \Phi_{+B}(\br,t)\\
0\\
0\\
\end{array}\right )
\end{equation}

\begin{equation}\label{psi-}
\Psi_-(\br,t)=
e^{-i\bK\cdot\br}\left( \begin{array}{c}
0\\
0\\
 \Phi_{-A}(\br,t)\\
 \Phi_{-B}(\br,t)\\
\end{array}\right )
\end{equation}
with the components

\begin{equation}
\Phi_{\zeta\alpha}(\br ,t)=\sum_{\lambda=\pm}\sum_{n,\ell}\langle \br |
\psi_{\zeta\alpha;\lambda n\ell}\rangle c_{\zeta;\lambda n\ell}(t)
\end{equation}
where $c_{\zeta;\lambda n\ell}(t)$ is the annihilation operator of an
electron in the state $|n,\ell\rangle$ of the $\lambda$ band with
valley index $\zeta$. The ${\rm K}$ valley ($\zeta=+$) part of the
single particle Green's function in reciprocal space can be written
as

\begin{equation}
G^0_+(\bq ,\omega)=\left(
\begin{array}{cc}
G^0_{+,AA}(\bq ,\omega) & G^0_{+,AB}(\bq ,\omega) \\
G^0_{+,BA}(\bq ,\omega) & G^0_{+,BB}(\bq ,\omega) \\
\end{array}\right),
\end{equation}
with matrix elements

\begin{equation}\label{G0komega}
G_{\zeta;\alpha\alpha^{\prime}}^0(\bk ,\omega)=
\sum_{\la}\sum_{n}\frac{f_{\zeta,\alpha\alpha^{\prime};\la n}(\bk +\zeta\bK) }{\omega -\la\xi_n +i\delta \sgn(\la\xi_n)},\\
\end{equation}
where $\la\xi_n=\la\epsilon_n-\epsilon_F$ is the energy difference
between the LL and the Fermi energy $\epsilon_F=\epsilon_{N_F}$,
which we choose in the conduction band ($\lambda=+$). Furthermore,
$\delta$ is a positive infinitesimal in the clean limit and the
matrix $f_{\la n}(\bq)$ has been derived in Ref. \onlinecite{RFG09}.
The expression for the Green's function in the $K'$ valley
($\zeta=-$) may be obtained with the help of
$G_{\zeta;\alpha\alpha'}^0= G_{-\zeta;\alpha'\alpha}^0$, and the
non-interacting polarization operator $\Pi^0(\bq,\omega)$ can be
calculated from Eq. (\ref{Pi0}). Taking into account both valleys,
the particle-hole polarization reads

\begin{equation}
i\Pi^0(\bq,\omega)
=i\left[\Pi^0_+(\bq,\omega)+\Pi^0_-(\bq,\omega)\right]=2i\Pi^0_+(\bq,\omega),
\end{equation}
where the last step indicates that one obtains equal contributions from both valleys. We may therefore restrict the
calculation to only one valley ($\zeta=+$), and take into account the twofold valley degeneracy by a simple factor
$g_v=2$.

The integration over the frequency integral yields then Eq. (\ref{Pi+}),
where the functions ${\cal
\overline{F}}_{nn^{\prime}}^{\lambda\lambda^{\prime}}(\bq)$ are given by
\begin{widetext}
\begin{equation}\label{FF}
{\cal
\overline{F}}_{nn^{\prime}}^{\lambda\lambda^{\prime}}(\bq)=\frac{e^{-l_B^2q^2/2}}{2\pi l_B^2}\left
( \frac{l_B^2q^2}{2}\right)^{\np-\nm}\left \{
\lambda 1_n^{*}1_{n'}^{*}\sqrt{\frac{(\nm-1)!}{(\np-1)!}}\left
[L_{\nm-1}^{\np-\nm}\left (\frac{l_B^2q^2}{2}\right )\right ]
+\lambda'2_n^{*}2_{n'}^{*}\sqrt{\frac{\nm!}{\np!}}\left [L_{\nm}^{\np-\nm}\left (\frac{l_B^2q^2}{2}\right )\right ]
\right \}^2.
\end{equation}
\end{widetext}

We define\cite{RFG09}
\begin{equation}
\Pi_{+,nn'}^{\,\,\,\la\lap}(\bq,\omega)=\frac{{\cal
\overline{F}}_{nn^{\prime}}^{\lambda\lambda^{\prime}}(\bq)}{\lambda\xn-\lambda^{\prime}\xnp
+\omega+i\delta }+(\omega^+\rightarrow-\omega^-)
\end{equation}
where $\omega^+\rightarrow\omega^-$ indicates the replacement $\omega+i\delta\rightarrow-\omega-i\delta$ and

\begin{eqnarray}
\Pi_{+,n}^{\la}(\bq,\omega)&=&\sum_{\lap}\sum_{n'=0}^{n-1}\Pi_{+,nn'}^{\,\,\,\la\lap}(\bq,\omega)\nonumber\\
&+&\sum_{\lap}\sum_{n'=n+1}^{N_c}\Pi_{+,nn'}^{\,\,\,\la\lap}(\bq,\omega)\nonumber\\
&+&\Pi_{+,nn}^{\la -\la}(\bq,\omega)
\end{eqnarray}
which verify
$\Pi_{+,n}^{\la}(\bq,\omega)=-\Pi_{+,n}^{-\la}(\bq,\omega)$. The
vacuum polarization, which accounts for the inter-band processes, is defined as

\begin{equation}
\Pi^{vac}_{+}(\bq,\omega)=-\sum_{n=1}^{N_c}\Pi_{+,n}^{\la=1}(\bq,\omega)
\end{equation}
where $N_c$ is a cutoff. Taking into account that, already
in the absence of magnetic field, the validity of the continuum
approximation is up to $\Lambda\sim t$, then
$\epsilon_{N_c}=(\vf/l_B)\sqrt{2N_c}\sim t$, which leads
to $N_c\sim 10^4/B[T]$, which is very high even for strong
magnetic fields. However, due to the fact that the separation
between LL in graphene decreases with $n$, it is always possible to
have {\it semiquantitative} good results from smaller values of
$N_c$.
\\

\section{Electron-hole pair momentum}\label{App:EHPM}

In this appendix, we relate the momentum $\bq$ of an
electron-hole pair to the distance between the guiding centers $\Delta \bR$ of the
electron and the hole. In classical mechanics, the cyclotron motion
of an electron leads to ${\pmb\pi}=e{\bf B}\times {\pmb \eta}$,
where ${\pmb\pi}$ is the gauge-invariant momentum and ${\pmb \eta}$
is the cyclotron coordinate. From Newton's equation with Lorentz's
force, it is obvious that the quantity ${\bf K}\equiv
{\pmb\pi}-e{\bf B}\times {\bf r}$ is a constant of the motion, where
${\bf r}$ is the electron position. This constant of the motion is
usually called the pseudo-momentum or generator of magnetic
translations.\cite{Y02} Defining the guiding center coordinate as
${\bf R}={\bf r}-{\pmb \eta}$, the pseudo-momentum reads ${\bf
K}=e{\bf R}\times {\bf B}$, which actually shows that, apart from a
conversion factor $eB=1/l_B^2$, the guiding center coordinate and
the pseudo-momentum correspond to the same constant of the motion.

Now consider an electron-hole pair, where the electron has a
pseudo-momentum ${\bf K'}$ and the hole a pseudo-momentum $-{\bf K}$
(corresponding to a removed electron of pseudo-momentum ${\bf K}$).
The pair has a momentum
\begin{equation}
{\bf q}\equiv {\bf K'}-{\bf K}=e \Delta{\bf R}\times {\bf B}
\end{equation}
and therefore $\Delta R=q l_B^2$, which is the sought after
relation.


\bibliography{BibliogrGrafeno}

\end{document}